\documentclass[fleqn,usenatbib,useAMS]{mnras}

\usepackage{graphicx}	% Including figure files
\usepackage{amsmath}	% Advanced maths commands
\usepackage{multicol}        % Multi-column entries in tables
\usepackage{bm}		% Bold maths symbols, including upright Greek
\usepackage{pdflscape}	% Landscape pages
\usepackage{csvsimple}
\newcommand{\zagaria}[0]{\textcolor{blue}{Zagaria et al. (in prep)}}

\usepackage{comment}
\usepackage[T1]{fontenc}
\usepackage{ae,aecompl}
\usepackage{pdfpages}
\usepackage{rotating}
\usepackage{newtxtext,newtxmath}
\usepackage[flushleft]{threeparttable}
\title{Observational Constraints on Evolution of Dust Disc Properties in Upper Scorpius}

\author[Pinilla, Sierra, et al.]
{
Paola Pinilla$^{1}$  \thanks{Contact e-mail: \href{mailto:p.pinilla@ucl.ac.uk}{p.pinilla@ucl.ac.uk}},
Anibal Sierra$^{1}$
\thanks{The first and second authors of this paper equally contributed to this work},
Nicolas T. Kurtovic$^{2}$,
Rossella Anania$^{3}$,
Sean Andrews$^{4}$,
John Carpenter$^{5}$,
\newauthor
Osmar Guerra-Alvarado$^{6}$,
Feng Long$^{7}$\thanks{NASA Hubble Fellowship Program Sagan Fellow},
Sebastian Marino$^{8}$,
Miguel Vioque$^{9}$,
Ke Zhang$^{10}$
\\ \\
$^{1}$ Mullard Space Science Laboratory, University College London, Holmbury St Mary, Dorking, Surrey RH5 6NT, UK\\
$^{2}$ Max Planck Institute for Extraterrestrial Physics, Giessenbachstrasse 1, D-85748 Garching, Germany\\
$^{3}$ Dipartimento di Fisica, Università degli Studi di Milano, Via Celoria 16, I-20133 Milano, Italy\\
$^{4}$ Center for Astrophysics, Harvard \& Smithsonian, 60 Garden St., Cambridge, MA 02138, USA\\
$^{5}$ Joint ALMA Observatory, Avenida Alonso de C\'ordova 3107, Vitacura, Santiago, Chile\\
$^{6}$ Leiden Observatory, Leiden University, PO Box 9513, 2300 RA Leiden, The Netherlands\\
$^{7}$ Lunar and Planetary Laboratory, the University of Arizona, Tucson, AZ 85721, USA\\
$^{8}$ University of Exeter, School of Physics and Astronomy, Astrophysics Group, Stocker Road, Exeter, EX4 4QL, UK \\
$^{9}$ European Southern Observatory, Karl-Schwarzschild-Str. 2, 85748 Garching bei München, Germany \\
$^{10}$ Department of Astronomy, University of Wisconsin-Madison, 475 N Charter St, Madison, WI 53706, USA.\\
}

\pubyear{{\the\year{}}}

\begin{document}
\label{firstpage}
\pagerange{\pageref{firstpage}--\pageref{lastpage}}
\maketitle

\begin{abstract}
Protoplanetary discs in the Upper Scorpius star-forming region are excellent laboratories to investigate late stages of planet formation. In this work, we analyse the morphology of the dust continuum emission of 121 discs from an ALMA Band 7 survey of the Upper Scorpius region. This analysis is done in the visibility plane, to measure the flux, geometry and characterise potential structures. We compare the results with state-of-the art gas and dust evolution models that include external photoevaporation, with mild values of the $F_{\rm{UV}}$ of 1-40\,$G_0$. From the visibility analysis, 52 of the 121 discs are resolved (43\%). From the resolved discs, 24 discs have structures and 28 remain as smooth discs at the mean resolution scale of $\sim$0.1$^{\prime \prime}$ ($\sim$ 14\,au).  Our results show no significant dust disc size evolution of the surviving discs in UpperSco when compared to discs in younger star-forming regions, such as Lupus. We find a strong, steeper-than-previously-reported correlation between dust disc size and disc millimeter continuum luminosity, in agreement with drift-dominated dust evolution models. We also find positive correlations between the dust disc mass vs. stellar mass and dust disc size vs. stellar mass. The slope of the dust disc size vs. stellar mass relationship is steeper compared to younger star forming regions. Additionally, we observe no significant correlation between dust disc properties and the environmental $F_{\rm{UV}}$, consistent with models predicting  that dust disc properties are primarily shaped by drift and dust traps. Our models predict that gas disc masses and sizes should be highly affected by the moderate $F_{\rm{UV}}$ values that Upper Scorpius discs experience in contrast to the dust. This highlights the need for deeper and higher-resolution ALMA observations of gas in these discs  exposed to mild external photoevaporation to further constrain their evolution and fate.
\end{abstract}


%%%%%%%%%%%%%%%%%%%%%%%%%%%%%%%%%%%%%%%%%%%%%%%%%%

%%%%%%%%%%%%%%%%% BODY OF PAPER %%%%%%%%%%%%%%%%%%

\section{Introduction}

The discovery and characterization of exoplanets over the past decades have revealed a remarkable diversity of planetary systems \citep{winn2015, lissauer2023, valencia2025}. A key challenge is to determine the processes that govern the chemical and physical evolution of protoplanetary discs—the birthplaces of planets—during planet formation. Demographic observations of protoplanetary discs that span a large range of ages and under different conditions are crucial to investigate the overall trends of disc properties, such as gas and dust mass/sizes. 

In the last decade, observations from the Atacama Large Millimeter Array (ALMA) have provided insights into the evolution of the  disc properties, in particular of the dust disc mass and size \citep[e.g.,][]{ansdell2016, baranfeld2016, cieza2019}, which has been used to compare with gas and dust evolution models to understand the physical process that dominate the evolution \citep[e.g.,][]{Rosotti2019b, pinilla2020, toci2021, zormpas2022}. Understanding the evolution of discs from early phases when they are still embedded (Class 0/I discs) to later stages, including the latest stages before the gas disc fully dispersed, is crucial to understand the overall picture of planet formation.

The Upper  Scorpius star-forming region (hereafter UpperSco) is often considered a standard reference for studying the later stages of the gas disc phase in protoplanetary disc evolution, as it has several populations with ages ranging from$\sim3$ and 10\,Myr old \citep[or even older e.g.,][]{pecaut2012, luhman2020, Armstrong2025}, which corresponds to the late stages of disc evolution \citep[e.g.,][]{fedele2010}. The disc millimetre fluxes are known to be lower in UpperSco in comparison to younger star forming regions \citep{baranfeld2016, pascucci2016, ansdell2017, carpenter2025}. In addition, the dust disc size seems to be smaller as well, as suggested by \cite{hendler2020}. 

However, only recently a comprehensive picture of the gas disc evolution has been done, albeit with only few discs. The large ALMA program AGE-PRO aimed to analyse the gas evolution of protoplanetary discs by observing  30 protoplanetary discs in three different star forming regions: Ophiuchus, Lupus, and UpperSco, representing three different stages of disc evolution. The  stellar spectral type in this sample was restricted to be between M3-K6, in order to have a narrow dependence with stellar type \citep{zhang_age-pro2025}. Two of the conclusions of the AGE-PRO collaboration are: (1) the median gas mass of the young discs ($< 1\,$Myr) is one order of magnitude higher than in the Lupus and UpperSco, while there is not a significant difference between the gas mass of the Lupus and UpperSco surviving discs \citep{trapman2025_mass}; (2) the median gas disc size obtained from the $^{12}$CO observations slightly increases from Lupus  to UpperSco \citep{trapman2025_size}.  Both of these results are interesting, partially because the discs in UpperSco are subject to mild external photoevaporation by $F_{\rm{UV}}$ radiation from OBA-type stars in the association, which can reduce the gas disc mass and truncate the disc efficiently \citep{anania2025_age-pro}. This challenges the comparison with younger star forming regions that have lower $F_{\rm{UV}}$ irradiation fields. 

In this study, we analyse continuum observations from the ALMA survey of UpperSco conducted by \cite{carpenter2025}. Compared to \cite{hendler2020}, we increase the sample from 22 discs to 121 discs, with a resolution that it is in average three times higher. The analysis is performed in the visibility domain to derive dust disc fluxes, sizes, and inclinations. For a subset of the UpperSco sample, we incorporate gas disc sizes derived by \zagaria\, to quantify dust–gas size ratios. We also compare the observations to disc models that include external photoevaporation, focusing on how FUV radiation fields typical of UpperSco influence  disc sizes and structures.

This paper is organised as follows. Section~\ref{sect:visi_modeling} explains the sample and procedure of the modelling of the data in the visibility plane. Section~\ref{sect:models} summarises the gas and dust evolution models of this work, in addition to initial conditions. Moreover, we explain in this section how different observables are calculated from the models in order to compare with ALMA data. Section~\ref{sect:results} includes all the results from the observations and models. Sections~\ref{sect:discussion} and~\ref{sect:conclusions} present the discussion and conclusions of our work, respectively.

%%%%%%%%%%%%%%%%%%%%%%
\section{Sample and Visibility Modelling} \label{sect:visi_modeling}
%%%%%%%%%%%%%%%%%%%%%%%

The data used in this work correspond to observations taken with ALMA in Band 7 (0.88mm) to detect the dust continuum and the CO J=3-2 \citep{carpenter2025}. The spectral windows of these observations were centred at 334.2, 336.1, 346.2, and 348.1\,GHz, with a bandwidth of 1.875\,GHz for each window. A total of 121 discs confirmed as members of UpperSco were detected in dust continuum emission with a signal-to-noise-ratio (SNR) $>$ 3, and 83 of these were also detected in CO with a SNR $>$ 5.  The resolution of these observations is between 0.1-0.3$^{\prime \prime}$, with a typical on-source integration time of 2.5 minutes, which provides a sensitivity of the dust continuum emission of 0.15\,mJy\,beam$^{-1}$.

We use \texttt{Galario} \citep{tazarri2018} to model the dust continuum visibilities of 121 discs in UpperSco and constrain their flux, size, and geometry. We assume axisymmetric discs, and an intensity radial profile described by a Gaussian function. The amplitude ($f_0$), width ($\sigma_r$), and radial peak position ($R_{\rm peak}$) of the Gaussian are free parameters of the fit. This model allows for the description of both smooth and ringed discs. In addition, the disc inclination (inc), position angle (PA), and offset from the phase centre (dRA, dDec) are also fitted for each disc, resulting in a total of seven free parameters.

This model is initially tested for all discs. However, the geometry and size (defined as the location where either 68\% or 90\% of the total flux  is enclosed) of some faint and compact discs are not well constrained as the corresponding fits do not  converge. In such cases, the brightness is modelled as a point source (a delta function $\delta$) with a total flux $F_{340 \rm GHz}$ and at position (dRA, dDec) relative to the phase centre. The Fourier transform of the point source is not computed with \texttt{Galario}, but directly from

\begin{eqnarray}
 \mathcal{F} [F_{340 \rm GHz}  \delta (\alpha_{\rm{RA}} - \textrm{dRA}, \, \beta_{\rm{Dec}} - \textrm{dDec})] &=& \\ \nonumber F_{340 \rm GHz} \exp(- 2 \pi i (u \, \textrm{dRA} + v \, \textrm{dDec})),
\end{eqnarray}

\noindent where $\alpha_{\rm{RA}}$ and $\beta_{\rm{DEC}}$ are the right ascension and declination coordinates, respectively, and $u$ and $v$ are the spatial frequencies in the Fourier domain (uv-distances). Since the discs are unresolved in this model, their geometry remains unconstrained, and their size upper limit is given by an angular resolution of 0.1$^{\prime \prime}$ ($14.5\,$au at an average distance of 145\,pc for the discs in UpperSco). A total of three free parameters (flux and offsets) are fitted for the unresolved discs.

In both cases (resolved and unresolved), the parameter space is explored using a Markov Chain Monte Carlo (MCMC), as implemented in the \texttt{Python} library \texttt{emcee} \citep{Foreman_2013}. The number of walkers per free parameter is set to 8. For resolved discs, the image models are generated with a resolution of $256 \times 256$ pixels and a pixel size of $0.02^{\prime \prime}$. The prior for each parameter follows a uniform distribution.
The number of steps is set to 30,000, which was found to be sufficient for parameter convergence. A final run of 10,000 steps is performed to estimate the parameters and their uncertainties. 

Tables~\ref{table:resolved_discs} and~\ref{table:unresolved_discs} present stellar and disc parameters obtained in this study, and in Section \ref{sec:Intensity_Profiles} we present and discuss the normalised intensity profiles of the resolved discs. We classify the resolved discs into two categories: structured discs and smooth discs. The former are defined as those in which the peak of the best-fitting Gaussian function is resolved ($R_{\rm peak} > 0.05^{\prime \prime}$), while in the latter, it remains unresolved ($R_{\rm peak} \leq 0.05^{\prime \prime}$). The value of $0.05^{\prime \prime}$ is chosen as an average value of the half of the beam size  of the observations for the sample \citep{carpenter2025}, and it is a conservative choice, as the typical resolution reached in the visibility  analysis is finer \citep{sierra2024}. It is important to note that in none of the clean images prominent multiple rings or more complex structures were clearly identified at the current resolution, justifying the choice of a single Gaussian, but with this approach we may still miss some structures as faint rings/gaps, azimuthal asymmetries, or small cavities.

\section{Models} \label{sect:models}
\subsection{Gas and Dust Evolution Models}

To perform the gas and dust evolution models of this work, we use the 1D code \texttt{Dustpy} \citep{stammler2022}, version 1.0.5. \texttt{Dustpy} is a publicly available code that follows the dynamics and growth of dust particles  simultaneously in the radial direction in protoplanetary discs, while the gas evolves by viscous evolution. To include external photoevaporation by ultraviolet (UV) radiation from nearby massive stars, we follow the prescription implemented by \cite{garate2024} in \texttt{Dustpy}. For this implementation, the gas surface density loss rate is calculated from the \texttt{FRIED} grid from \cite{haworth2018}, and for the loss of dust particles that are entrained by the photoevaporative wind we use the models by \cite{sellek2020}. We used an updated version of the \texttt{FRIED} grid \citep{haworth2023}, as implemented in \texttt{Dustpy} in \cite{anania2025_age-pro}.

Based on the properties of the UpperSco discs \citep{carpenter2025}, we performed simulations for two different types of stars: a solar type star with a stellar mass of $1.0$\,$M_\odot$  and a luminosity of $1.0$\,$L_\odot$; and an M-dwarf with a stellar mass of $0.3$\,$M_\odot$ and a luminosity of $0.3$\,$L_\odot$. In the sample of \cite{carpenter2025}, most of the discs are around M3-M5 stars, so our second choice is more representative when we compare to the data. 

The gas surface density is assumed to be a power law with an exponential cut-off \citep{lyndenbell1974}, that is,

\begin{equation}
\Sigma_g(r) = \Sigma_0 \left(\frac{r}{r_c} \right)^{-\gamma} \ \exp\left[ -\left( \frac{r}{r_c}\right)^{2-\gamma} \right],
 \label{eq:sigma_gas}
\end{equation}

\noindent and we assume the cut-off radius $r_c$ to be 80\,au, and $\gamma=1$. The initial disc mass is $M_{\rm{disc}}=0.05\,M_{\star}$, which sets the value of $\Sigma_0$ for each of the stellar masses assumed in this work. Although the value of $r_c$ is large compared to values obtained from observations \citep{trapman2025_size}, \cite{kurtovic2025_age-pro} and \cite{anania2025_age-pro} showed that varying $r_c$ has low influence on the gas and dust evolution over million year timescales. We chose $r_c=80$\,au to allow that the assumed traps are able to keep dust particles in the outer disc over million years of evolution, as traps assumed outside $r_c$ trap very little dust available in the outer disc \citep{garate2024}.

For the disc temperature, we assume only the stellar irradiation \citep{kenyon1987}. The radial grid in our simulations is from 1 to 300\,au with 150 grid cells, which is logarithmically spaced. For some of our simulations, we perform resolution tests with a double number of cells, resulting in no differences on the final  disc properties and evolution. For the dust grain size distribution $n(a)$, we assume that it is initially distributed as a power law as $n(a)\propto a^{-3.5}$ (with $a$ being the grain size) from 0.5-1\,$\mu$m. The exponent of -3.5 is assumed as the value from the size distribution of interstellar grains \citep{mathis1977}. The grain size grid is logarithmically spaced, and it spans from  0.5\,$\mu$m to 25\,cm, with  cells. The initial dust-to-gas ratio is assumed to be as the interstellar medium, with a value of 1/100. The grains in the simulations are allowed to grow and fragment. Particles fragment when they reach a threshold in their relative velocities, which is called the fragmentation velocity, and it is assumed in this work to be constant and equal to 10\,m\,s$^{-1}$, as typically assumed for water-ice particles \citep[e.g.,][]{Gundlach2015}

We explored four values of $F_{\rm{UV}}$, that is, 1, 10, 20, and 40\,$G_0$; based on the calculations in \zagaria\,for the sources used in this work in UpperSco. We  turn-on the effect of external photoevaporation at 1\,Myr of evolution, as we assume that at early ages the discs could have been shielded by the envelope  \citep{qiao2022}. \cite{anania2025_age-pro} demonstrated that the assumption of turning-on external photoevaporation later in the disc evolution does not change the results at longer times, because in this case the disc properties rapidly evolve to the same values as if external photoevaporation would be active from the beginning of the simulation. 

We include the presence of pressure bumps in our models, by adding a perturbation to the viscous $\alpha_{\rm{visc}}$ profile, which creates a gap-like structure in the gas surface density profile, as in \cite{stadler2022}, \cite{garate2024} and \cite{kurtovic2025_age-pro}. The perturbation in the $\alpha$ profile has the shape of a Gaussian bump, that is,
\begin{equation} \label{eq_alpha_bump}
    \alpha_{\rm{visc}}(r) = \alpha_0 \left(1 + A \exp\left(-\frac{\left(r - r_\textrm{gap}\right)^2}{2w_\textrm{gap}^2}   \right)\right),
\end{equation}

\noindent where $\alpha_0$ is the global viscous disc parameter \citep{shakura1973} that is assumed to be $10^{-3}$ for all simulations. $A$ sets the strength of the pressure bumps, and it is taken to be $A=0,1,4,50$, where $A=1, 4$ represents weak and strong pressure bumps, respectively, as implemented in \cite{pinilla2020}. In the case of $A=1$ and $A=4$, there are three bumps or gaps considered simultaneously in the disc and centred at $r_{\rm{gap}}= 10, 40, 70$\,au, all of them with a width $w_{\rm{gap}}$ equal to the local pressure scale height. The number and location of the traps is motivated by  observations of protoplanetary discs. In the compilation of discs by \cite{bae2023} most of the observed rings have been found between $20-80$\,au. The case of $A=50$ is assumed to mimic a very deep gap as observed in discs with large cavities, the well-known transition discs (hereafter transition disc-type trap). In this case, only one gap is considered at $r_{\rm{gap}}$= 40\,au, with a width equal to three scale heights. We run the simulations from 0-10\,Myrs, saving a total of 12 snapshots at $t_{\rm{snapshots}}=[0.001, 0.005, 0.01, 0.05, 0.1, 0.5, 1, 2, 3, 5, 7, 10]\,$Myr. All the model parameters are summarised in Table~\ref{Table_Parameters}.

\begin{table}
\caption{Parameter space of the gas and dust evolution models}
\label{Table_Parameters}
\begin{tabular}{l l l }
 \hline \hline
Quantity/Unit & Description & Value  \\
\hline
$M_\star$[$M_\odot$]& Stellar mass & 0.3, 1.0  \\
$T_\star$[K] & Stellar effective temperature& 3000, 5772\\
$L_\star$[$L_\odot$] & Stellar luminosity& 0.3, 1.0\\
$r_{\rm{in}}$ [au] & Inner radial boundary& 1.0 \\
$r_{\rm{out}}$ [au] & Outer radial boundary& 300 \\
$r_c$ [au]&Cut-off radius& 80\\
$n_r$&Number of radial cells&150\\
$M_{\rm{disc}}$ [$M_\star$]&Disc mass&0.05\\
$\alpha_0$&Disc viscosity&$10^{-3}$\\
$v_f$ [m\,s$^{-1}$]&Fragmentation velocity&10\\
$\rho_s$[g\,cm$^{-3}$] & Grain material density & 1.7\\
$A$ & Amplitude of pressure bumps & 0, 1, 4, 50\\
$r_{\rm{gap}}$ [au] & Location of gaps for $A=1,4$ & 10, 40, 70\\
$r_{\rm{gap}}$ [au] & Location of gap for $A=50$ & 40\\
$F_{\rm{UV}}$ [$G_0$] & External UV Flux & 1, 10, 20, 40\\

\hline
\end{tabular}
\vspace{-3.0mm}
\end{table}

\begin{figure*}
    \centering
    \includegraphics[width=\textwidth]{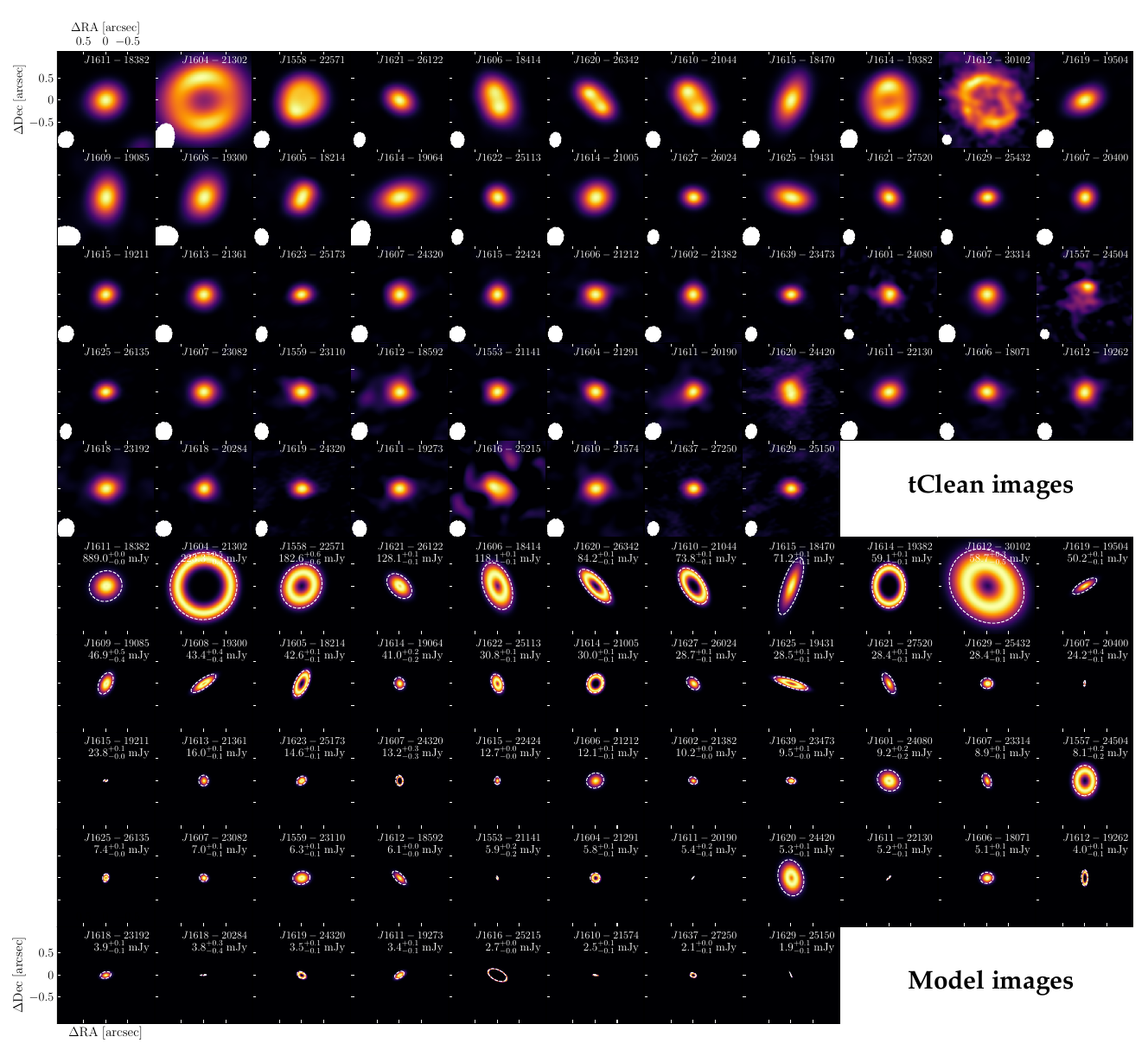}
    \caption{Gallery of the ALMA images (top panels) and the corresponding visibility modelling images (bottom panels) of the discs that are resolved, organised by millimetre brightness. Each image is normalised to the peak and shown with a linear colour stretch. In the visibility modelling images, the dashed-white lines represent $R_{90}$, and the  total flux is given in each panel.}
    \label{fig:rad_profile}
\end{figure*}

\begin{figure*}
    \centering
    \includegraphics[width=\textwidth]{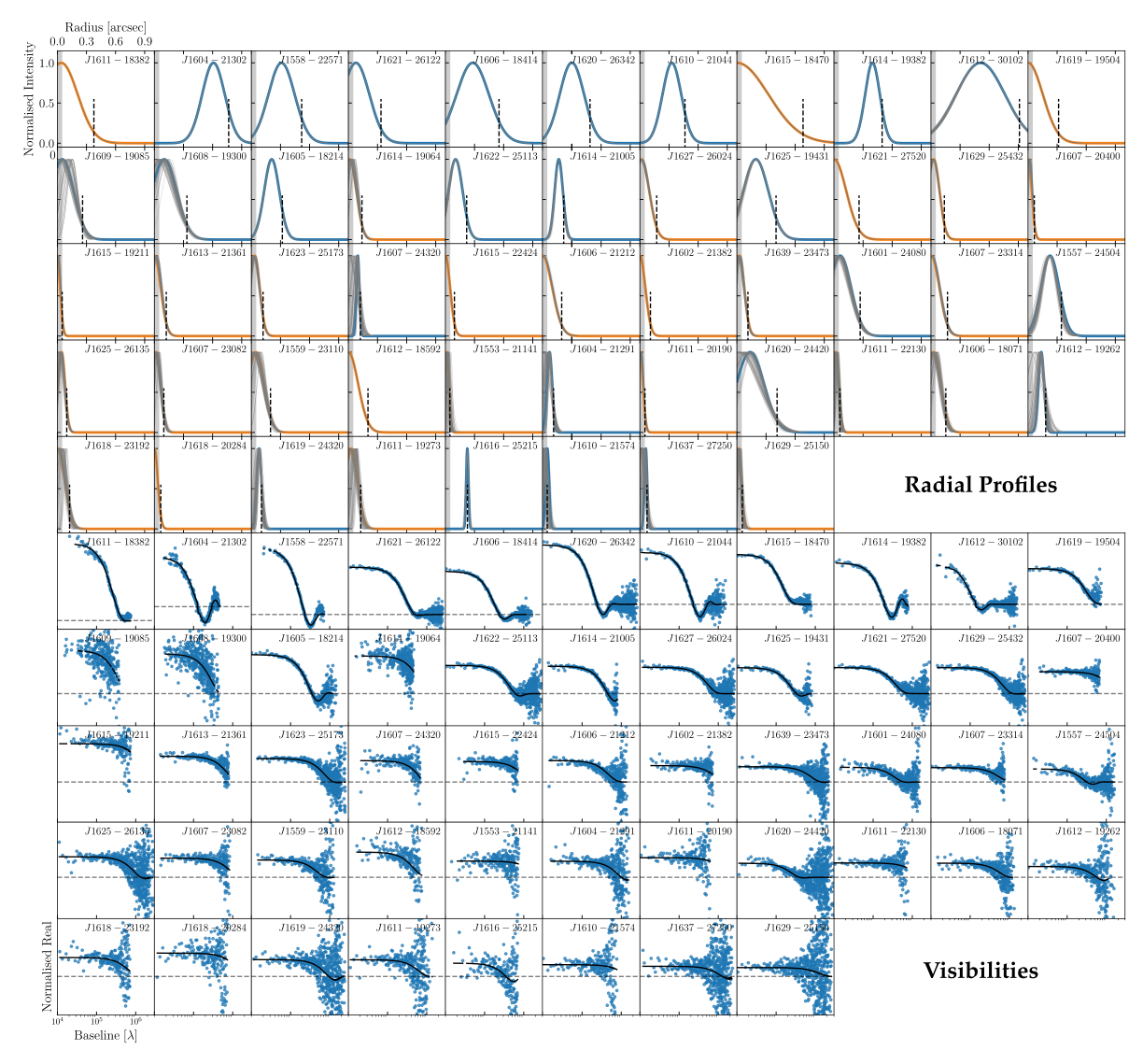}
    \caption{Upper panels: Normalised intensity radial profiles (orange for smooth discs and blue for structured discs)  obtained from the visibility modelling for the resolved discs in our sample. The dashed vertical lines indicate the position of $R_{90}$. The gray shaded area indicates the region within $0.05^{\prime \prime}$ (1/2 beam size). We over-plot the last 250 models for each fit to show the typical uncertainties of the fits. Bottom panels: Real part of the visibilities, binned at 3k$\lambda$ (blue points), and the best fit model (black lines) for all the resolved discs in the sample.}
    \label{fig:rad_profile2}
\end{figure*}

\subsection{Synthetic observable properties} \label{sect:synthetic_obs_calculations}

To compare with the ALMA observations, we calculate different observable properties using the dust size distribution obtained for each snapshot of the \texttt{DustPy} simulations. These dust distributions are used to calculate the optical depth at each radius $\tau_\nu(r)$ with $\nu=340$\,GHz as

\begin{equation}
    \tau_\nu (r) = \frac{\sum_a \kappa_\nu(a)\, \sigma (a, r)}{\cos i},
\end{equation} 

\noindent where $\sigma (a, r)$ is the vertically integrated dust surface density and $\kappa_\nu(a)$ are the dust opacities, for which the dust composition is assumed to be as in \cite{ricci2010}, but for compact grains. The disc inclination is assumed to be $i=45^{\circ}$.

\begin{figure*}
    \centering
    \includegraphics[width=0.9\textwidth]{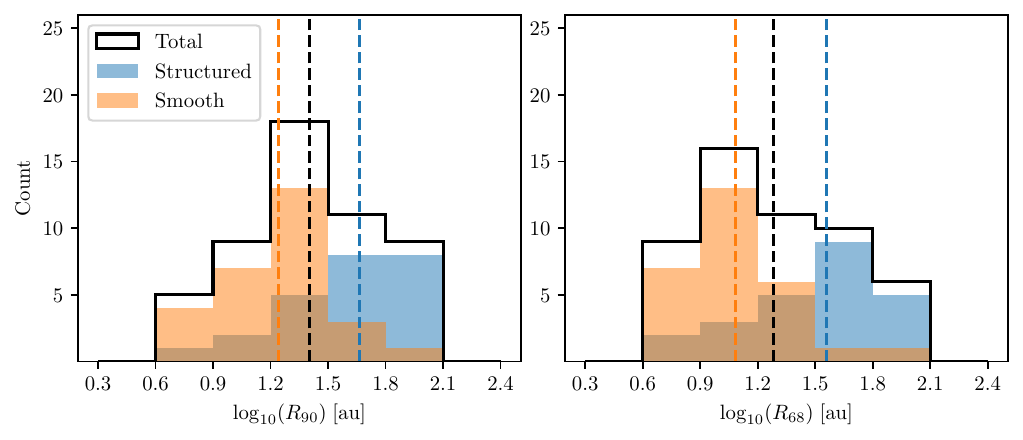}
    \caption{Distribution of $R_{90}$ (left) and $R_{68}$ (right) inferred from dust continuum visibility modelling for the resolved discs with structures (blue) smooth discs (orange), and all resolved discs (in black). The vertical line indicates the median value of each distribution. }
    \label{fig:Sizes}
\end{figure*}

\begin{figure}
    \centering
    \includegraphics[width=0.9\linewidth]{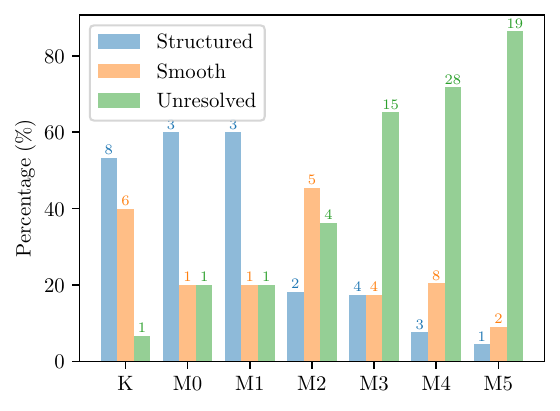}
    \caption{Distribution of unresolved, structured, and smooth discs in UpperSco based on the spectral type of the hosting star. The number of discs in each category is shown on the top of each bar.}
    \label{fig:Spectral-type}
\end{figure}

For the calculation of the flux at 340\,GHz (880\,mm), we assume for simplicity only thermal emission from the dust, so the intensity as a function of radius is given by,

\begin{equation} \label{eq_Flux_OpticallyThin}
    I_\nu (r) = B_\nu(T(r)) \left[1 - \exp(-\tau_\nu(r)) \right], 
\end{equation}

\noindent where $B_\nu(T)$ is the Planck function at a  temperature of $T(r)$. While in the models, the gas and dust have the same temperature ($T_g$), for the calculation of the disc observables, we assume there is a background temperature of $T_\textrm{b} =20$\,K, to account for the irradiation from other stars in the region as in \cite{garate2024}, this means that 

\begin{equation}
    T(r) = \sqrt[4]{T_g(r)^4 + T_\textrm{b}^4}.
    \label{eq:temp}
\end{equation}

The intensity profile is convolved with a Gaussian beam of size 0.1$^{\prime \prime}$, to account for a similar resolution of the UpperSco observations. The total flux is calculated as $F_\nu = \int I_\nu\, d{\Omega}$, where $d{\Omega}$ is the solid angle covered by the disc in the sky, assuming a distance of 145\,pc.

The dust disc radius is defined as the location that encloses 90\% of the flux ($R_{90}$). For this, we assume a sensitivity 0.15\,mJy\,beam$^{-1}$, similar to the value of the observations \citep{carpenter2025}. This means that in our calculations, any value of the flux lower than three times this sensitivity threshold is assumed as a non detection.  

\begin{figure*}
    \centering
    \includegraphics[width=\textwidth]{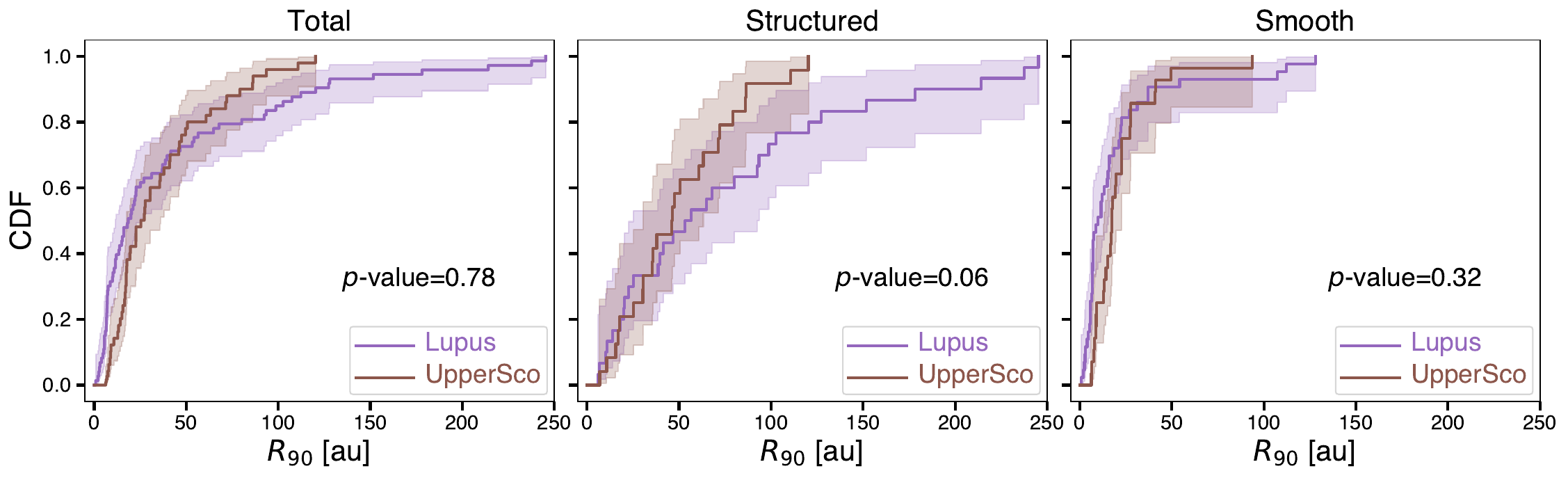}
    \caption{Cumulative distribution functions (CDFs) of the dust continuum disc radii ($R_{90}$) for the Lupus sample \citep{guerra2025} and the UpperSco sample of this study. Left to right panels correspond to the: total (including upper limits for the case of UpperSco), structured and smooth discs sample in each region. The panels show the $p-$ values, which are higher than $1\%$ in all cases, implying  that the two distributions remain undistinguishable.}
    \label{fig:CDFS}
\end{figure*}

In addition, we calculate the gas disc size from the models as the location that encloses  90\% of $^{12}$CO line flux (hereafter, $R_{\mathrm{CO}}$). To obtain this location, we assume the analytical prescriptions in \cite{trapman2023}. They used  thermochemical models to find an empirical correlation between the gas column density at the observed gas disc size $R_{\mathrm{CO}}$ and the mass of the disc, which does not significantly depend on other disc parameters. This column density is

\begin{equation}
        N_{\mathrm{gas}}(R_{\mathrm{CO}}) \approx 10^{21.27 - 0.53 \log(L_{\star})} \left( \frac{M_{\mathrm{gas}}}{M_{\odot}} \right)^{0.3 - 0.08 \log(L_{\star})},
        \label{n_crit}
\end{equation}

\noindent where $L_\star$ is the stellar luminosity and $M_{\rm{gas}}$ is the gas disc mass. The $R_{\mathrm{CO}}$ value is then calculated when $ \Sigma_g (R_{\mathrm{CO}}) = N_{\mathrm{gas}}(R_{\mathrm{CO}}) \mu_{\mathrm{gas}}$, with $\mu_{\mathrm{gas}}$ being 2.3\,$\mu_H$ ($\mu_H$ is the mass of atomic hydrogen). It is worth noticing that the background temperature added in Eq.~\ref{eq:temp} does not affect the calculation of $N_{\mathrm{gas}}(R_{\mathrm{CO}})$ in Eq.~\ref{n_crit}.

%%%%%%%%%%%
\section{Results} \label{sect:results}
\subsection{Intensity profiles and dust disc size}
\label{sec:Intensity_Profiles}
%%%%%%%%%%%%
From the  analysis in the visibility plane of the entire sample, we find that 24 discs ($\sim$20\%) are resolved and structured, 28 ($\sim23$\%) are resolved and remain smooth at the current resolution; and 69 ($\sim$57\%) are unresolved. Figure~\ref{fig:rad_profile} shows 
the ALMA images (top panels) and the  the model images from  the visibility fitting (bottom panels) for all the resolved discs in the sample. The total flux of each disc is given in the individual panels of the images from  the visibility fitting. In addition, Fig.~\ref{fig:rad_profile2} shows the best intensity profile (normalised to the peak) for each of the discs that are resolved (top panels). The vertical lines in this figure  correspond to the position of $R_{90}$, which is the location that encloses 90\% of the total continuum flux. The bottom panels of Fig.~\ref{fig:rad_profile2} are the real part of the binned visibilities, with the best fit model overplotted. In this paper, we  also use $R_{68}$, corresponding to the location that encloses 68\% of the total flux, commonly used in previous papers \citep[e.g.,][]{tripathi2017, andrews2018, hendler2020, kurtovic2021, vioque2025_age-pro}. Figure~\ref{fig:Comparison_Literature} in the Appendix shows the fluxes, dust disc sizes, and inclinations of the fit in comparison with the values reported in \cite{carpenter2025} and \zagaria, showing in general good agreements. It is worth noting that for all the figures in this work, we use the values obtained from the visibility modelling for the  dust disc sizes only for the resolved discs. For the unresolved discs, the disc size upper limit is given by an average of the beam size in the image plane. For the fluxes reported in figures, we use the values of the visibility fitting for the resolved discs, while for the unresolved discs we use  the values from \cite{carpenter2025}, as their typical uncertainties are lower. This is because in our visibility fitting, we do not impose a zero offset,  so the point source model can wander to noise peaks. The values for the stellar and disc properties obtained in this work are summarised in Tables~\ref{table:resolved_discs} and ~\ref{table:unresolved_discs}.

Figure~\ref{fig:Sizes} shows the distribution of $R_{90}$ (left) and $R_{68}$ (right) inferred from dust continuum visibility modelling for discs with structures (blue), smooth discs (orange), and all resolved discs (black). The discs that are unresolved are not included. The vertical line indicates the median value of each distribution, which are summarised in Table~\ref{Table:R90_R68}. We compare the cumulative distribution function of smooth vs. the structured discs using the  two-sample Kolmogorov-Smirnov test (\texttt{K-S} test), and we find that these two distributions are significantly different. 
Figure~\ref{fig:Spectral-type} shows the distribution of unresolved, structured, and smooth discs according to spectral type. This histogram  shows that unresolved discs are more commonly found around later-type stars. 

The rate of discs that have been classified as structured with a comparable resolution ($\sim$0.1$^{\prime \prime}$) is similar than in Taurus \citep[30\%,][]{long2019}, which is a younger star-forming region ($\sim1-3$\,Myr) than UpperSco. The analysis in \cite{long2019} was limited to 30 discs, which stellar types are earlier than M2. As it is shown in Fig.~\ref{fig:Spectral-type} the rate of structured discs in our UpperSco sample decreases for late M-dwarfs. For discs around K-, M0-, and M1- stars, the structured discs rate ranges between $\sim$50-60\%, and it decreases to 20\% for M2 and M3.  A follow-up study of the same Taurus sample from \cite{long2019}  revealed additional compact or smooth discs with structures at $\sim$0.1$^{\prime \prime}$ resolution, increasing the detection rate to $\sim$56\% \citep{zhang_s2023}, which is similar to our rates for M1-K stars. 

The discs in UpperSco analysed by the AGE-PRO collaboration are a subset of the analysed sample in this work, but observed at 0.2-0.3$^{\prime \prime}$ resolution and with  much higher sensitivity \citep{agurto_age-pro2025}. When restricting the calculation of the median value of $R_{90}$ and $R_{68}$ to the same spectral type as AGE-PRO (stellar types between M3-K6), we found $\sim$27\,au and $\sim$19\,au, respectively, which are similar to the median values found in the AGE-PRO sample \citep[][]{vioque2025_age-pro}.

\cite{guerra2025} recently analysed a total of 73 discs in Lupus, covering a spectral type between M5-K2, observed with ALMA at a resolution of 0.03-0.04$^{\prime \prime}$, and demonstrated that around 67\% of the discs in Lupus are smaller than 30\,au (most of them around M-type stars, 73\%), and within that group the rate of disc with structures is similar to our rate in UpperSco (28\% in Lupus), while for large discs ($>30$\,au) the large majority of discs ($\sim$80\%) in Lupus have structures. 
We compare the distribution of the dust continuum disc radii ($R_{90}$) in Lupus vs. UpperSco in Fig.~\ref{fig:CDFS}, which shows the  cumulative distribution functions (CDFs) when  using  the total, structured, and smooth disc sample using the Kaplan-Meier estimator \citep[as implemented in \texttt{lifelines},][]{Davidson-Pilon2019}. The $p-$value obtained from the \texttt{logrank} test from \texttt{lifelines}, which includes upper limits for the unresolved discs in the total sample is reported Fig.~\ref{fig:CDFS}. These values are higher than $1\%$ in all cases, implying  that the distributions between Lupus and UpperSco remain undistinguishable and that there is no significant evolution of $R_{90}$ between the two regions that mark different disc evolutionary stages. The median values for Lupus are summarised in Table~\ref{Table:R90_R68}.

\begin{figure*}
    \centering
    \includegraphics[width=0.9\linewidth]{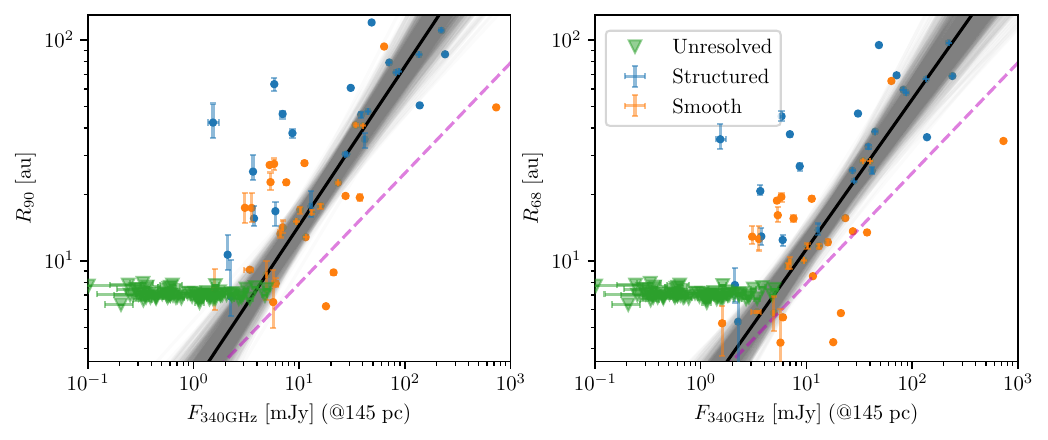}
    \caption{$R_{90}$ (left panel) and $R_{68}$ (right panel) vs. the dust continuum flux at Band 7 normalized at 145 pc. Blue and orange dots indicate discs with and without structures, respectively. The best fit from MCMC linear regression is shown as a black solid line. 1000 random posterior fits are plotted using gray lines to visualize the fit uncertainty. We overplot  a line corresponding to an optical depth of $\tau=\infty$ (dashed  magenta line), assuming a dust temperature of 20\,K.}
    \label{fig:Flux-Size}
\end{figure*}

\begin{table}
\centering
\caption{Median values of dust continuum disc radii, $R_{90}$.}
\label{Table:R90_R68}
\begin{tabular}{l|l|l}
\hline 
& UpperSco &  Lupus  \\
& $R_{90}$ [au] & $R_{90}$ [au]   \\
\hline 
Total & $25.4^{+10.2}_{-7.7}$ &$18.5^{+6.6}_{-6.9}$\\\\
Smooth &$17.4^{+5.2}_{-4.1}$ &$9.9^{+5.4}_{-3.0}$ \\\\
Structured &$46.3^{+16.9}_{-15.9}$ &$53.3^{+40.1}_{-28.2}$ \\
\hline
\end{tabular} \\
\begin{tablenotes}
\item \small{\textbf{Notes.} The errors correspond to the 95\% confidence regions (or $2\sigma$) using the Kaplan-Meier estimator. Upper limits are included for the discs in the total sample of UpperSco and that are not resolved in this study. The Lupus values are calculated from \cite{guerra2025}. }
\end{tablenotes} 
\end{table}

\subsection{Dust disc size vs. millimetre flux} \label{sect:Lmm-R90}

Figure~\ref{fig:Flux-Size} shows the $R_{90}$ (left panel) and the $R_{68}$ (right panel) radius vs. the dust continuum flux at Band 7 (340\,GHz or $\sim$0.88\,mm) normalised at 145 pc. We also show in this figure the line corresponding to an optical depth of $\tau\gg1$ (dashed  magenta line), assuming a dust temperature of 20\,K. Optically thinner discs lie on the left side of this line.

To quantify a potential correlation between these quantities, we perform a Bayesian linear regression using \texttt{linmix} \citep{kelly2007}, such that $\log(R_{90}) = \alpha + \beta \log(F_{340\rm{GHz}})$, and similarly for $R_{68}$. For this fit, we  include the unresolved discs (or upper limits in this fit).  The values for the slope and the intercept of this fit are summarised in Table~\ref{Table:summary_fits}.

\begin{table*}
\centering
\caption{Summary of the fits}
\label{Table:summary_fits}
\begin{tabular}{l|l l l l l l}
\hline 
Sample & $x$ & $y$ &$\alpha$&$\beta$& $\sigma$ & $\hat{\rho}$, CI$_\rho$\\
\hline \hline
Total & $\log(F_{340\rm{GHz}})/$mJy & $\log(R_{90})/$au & $0.44^{+0.09}_{-0.10}$ &$0.72^{+0.08}_{-0.07}$ & $0.10^{+0.03}_{-0.02}$ & $0.87,  (0.84, 0.89)$\\
& $\log(F_{340\rm{GHz}})$/mJy & $\log(R_{68})/$au & $0.36^{+0.07}_{-0.06}$&$0.68^{+0.07}_{-0.06}$&$0.10^{+0.03}_{-0.02}$&$0.86,  (0.82, 0.88)$ \\
& $\log(M_\star)/M_{\odot}$ & $\log(F_{340\rm{GHz}})/$mJy & $1.71^{+0.15}_{-0.15}$&$1.83^{+0.22}_{-0.22}$&$0.21^{+0.05}_{-0.04}$&$0.78,  (0.72, 0.83)$\\ 
& $\log(M_\star)/M_{\odot}$ & $\log(M_{\rm{dust}})/M_{\oplus}$ &
$1.14^{+0.15}_{-0.15}$&$1.83^{+0.22}_{-0.22}$&$0.21^{+0.05}_{-0.04}$&$0.78,  (0.72, 0.83)$\\
& $\log(M_\star)/M_{\odot}$ & $\log(R_{90})/$au & $1.91^{+0.14}_{-0.13}$ &$1.69^{+0.28}_{-0.24}$ & $0.16^{+0.08}_{-0.05}$ & $0.81,  (0.74, 0.87)$\\
\hline
Structured  & $\log(F_{340\rm{GHz}})/$mJy & $\log(R_{90})/$au & $1.13^{+0.10}_{-0.11}$&$0.36^{+0.07}_{-0.07}$&$0.04^{+0.02}_{-0.01}$&$0.78,  (0.67, 0.86)$ \\
& $\log(F_{340\rm{GHz}})$/mJy & $\log(R_{68})/$au & $1.00^{+0.10}_{-0.11}$&$0.38^{+0.07}_{-0.07}$&$0.05^{+0.02}_{-0.01}$&$0.76,  (0.65, 0.85)$\\
\hline
Smooth and & $\log(F_{340\rm{GHz}})/$mJy & $\log(R_{90})/$au & $0.21^{+0.13}_{-0.15}$&$0.82^{+0.13}_{-0.11}$&$0.13^{+0.06}_{-0.04}$&$0.83,  (0.78, 0.87)$ \\
unresolved& $\log(F_{340\rm{GHz}})$/mJy & $\log(R_{68})/$au & $0.21^{+0.10}_{-0.14}$&$0.71^{+0.11}_{-0.09}$&$0.10^{+0.04}_{-0.03}$&$0.82,  (0.77, 0.87)$\\
\hline
Only M3-K6 & $\log(F_{340\rm{GHz}})/$mJy & $\log(R_{90})/$au &
$0.40^{+0.16}_{-0.18}$&$0.81^{+0.14}_{-0.12}$&$0.12^{+0.05}_{-0.03}$&$0.87,  (0.81, 0.90)$ \\
& $\log(F_{340\rm{GHz}})$/mJy & $\log(R_{68})/$au & 
$0.27^{+0.15}_{-0.18}$&$0.79^{+0.14}_{-0.12}$&$0.10^{+0.04}_{-0.03}$&$0.88,  (0.83, 0.91)$ \\
\hline
\end{tabular} \\
\begin{tablenotes}
\item \small{\textbf{Notes.} Summary of the fits of this study using \texttt{linmix}, assuming $y=\alpha+\beta x$. Upper limits are included when unresolved discs are part of the sample. The reported values correspond to the median, and the uncertainties represent the 68\%\ confidence interval. The variance $\sigma$ or regression intrinsic scatter as well as  the correlation coefficient ($\rho$) are reported with 68\%\ of their confidence intervals.}
\end{tablenotes} 
\end{table*}

In addition, we calculate the variance or the regression intrinsic scatter ($\sigma$) and the correlation
coefficient ($\rho$), which are also reported in Table~\ref{Table:summary_fits}.  Both correlations are steeper and stronger (that is, a  higher correlation coefficient) than previously reported in \cite{hendler2020} for UpperSco. This correlation can partially  be the result of  the optical depth of discs, as the $\tau\gg1$ line in Fig.~\ref{fig:Flux-Size} suggests, which is discussed in more details in Sect.~\ref{sect:discussion}.

In \cite{hendler2020}, the value of the slope for the dust disc size vs. luminosity relationship using a sample of 22 discs in UpperSco is much flatter (specifically: $R_{68}\propto L_{\rm{mm}}^{0.22}$) than we find in this study ($R_{68}\propto L_{\rm{mm}}^{0.68}$).  \cite{hendler2020} found a steeper correlation for younger star forming regions, including Ophiuchus, Taurus, Lupus, and Chamaeleon I, with slopes between 0.60-0.40. They also covered a large span of stellar masses between $\sim$0.08-4\,$M_\odot$. In \cite{andrews2018}, an analysis of  ALMA and SMA observations of 105 discs in Lupus, Taurus and Ophiuchus gave a relationship such that  $R_{68}\propto L_{\rm{mm}}^{0.49}$.

In the analysis of \cite{guerra2025} of discs in Lupus, the slope of this correlation is $0.61\pm 0.06$,  similar to the value reported in \cite{tripathi2017, andrews2018, hendler2020} for Lupus, which within the uncertainties is similar to the correlation found for UpperSco in this work. Therefore, we suggest that the dust disc size vs. millimetre flux relationship does not show any evolutionary change when comparing younger star forming regions to UpperSco. 

\begin{figure}
    \centering
    \includegraphics[width=0.9\linewidth]{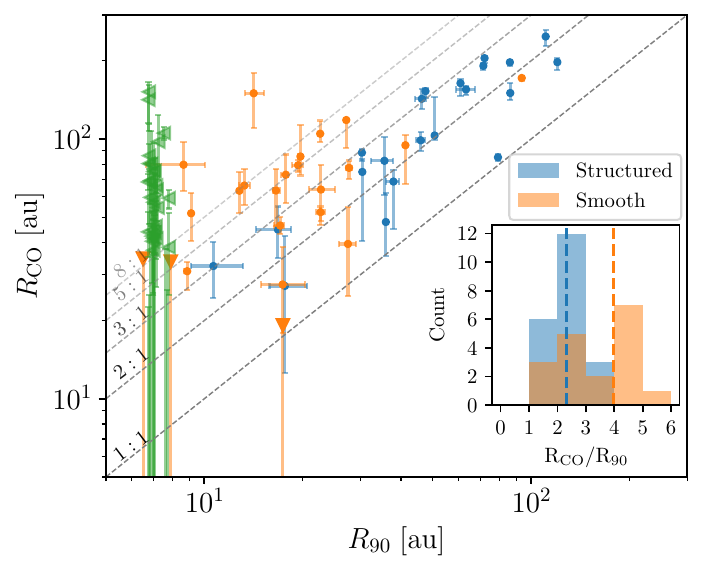}\\
    \includegraphics[width=0.85\linewidth]{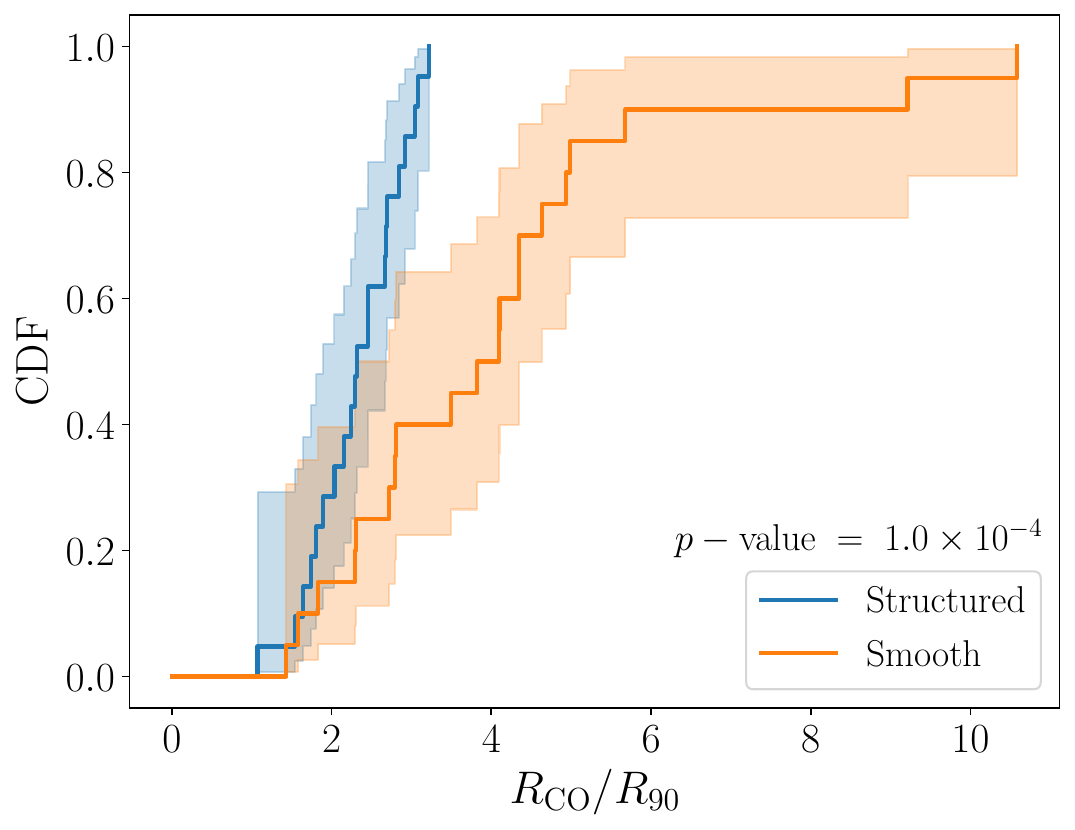}\\
    \caption{Top panel: $R_{CO}$ vs. $R_{90}$ radius from dust continuum emission. The dashed lines indicate  CO-to-continuum radius ratios equal to 1, 2, 3, 5, 8. 
    The inset panel shows the histogram of the CO-to-continuum radius ratio for structured (blue) and smooth (orange) discs. Bottom panel: CDFs of the $R_{\rm{CO}}$/$R_{90}$ for the structured vs. smooth discs in the UpperSco sample of this study. }
    \label{fig:RCO-Rcont}
\end{figure}

\cite{vioque2025_age-pro} performed an analysis of this correlation for the discs of the AGE-PRO sample,  including 10 discs in UpperSco, which are part of the sample analysed in this work. They did not find a correlation between the dust disc size and disc luminosity. They also used the data from \cite{hendler2020} sample and limited it to the AGE-PRO spectral type range and found no correlation. 
The correlation is found in our analysis when we restrict the stellar type as in the AGE-PRO sample, with similar values than the  whole sample of the slope and intercept (Table~\ref{Table:summary_fits}), both of them with  correlation coefficients that indicate a positive correlation. Therefore, when restricting the stellar type M3-K6, this correlation only becomes clear in UpperSco with the full sample analysed in this work.

\begin{figure*}
    \centering
    \includegraphics[width=0.9\textwidth]{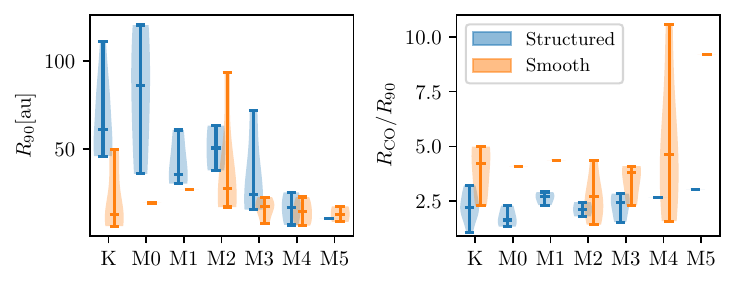}\\
    \includegraphics[width=0.9\textwidth]{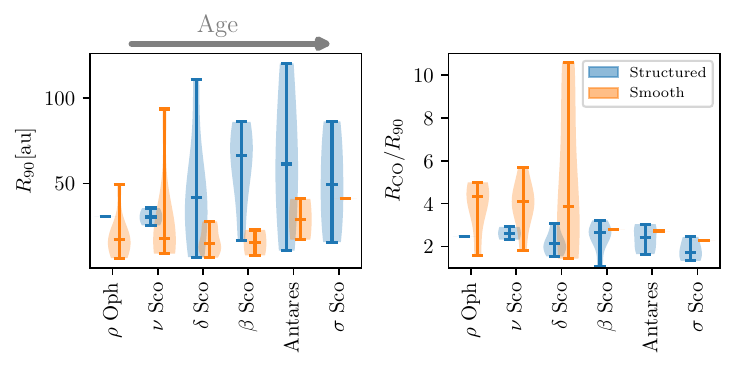}
    \caption{Distribution of $R_{90}$, $R_{\rm{CO}}$, and $R_{\rm{CO}}$/$R_{90}$ for the structured (blue) and smooth (orange) discs based on the spectral type of the central star (top panels), and sub-region that they belong, organised by age (bottom panels). Some entries do not have a  distribution because there are few number of discs in that bin.}
    \label{fig:R90_SpT}
\end{figure*}

In addition,  we also perform the fit of the smooth and unresolved discs separately from the structured discs, finding a steeper slope for the smooth/unresolved vs. structured discs (Table~\ref{Table:summary_fits}). We note that the large difference in the slope is driven by the upper limits (unresolved discs). When the upper limits are not included in the fit, the smooth and structured discs have a similar slope in the dust disc size vs. millimetre flux relationship.

%%%%%
\subsection{Gas vs. Dust Disc Size}
%%%%%
\zagaria \, measured gas disc sizes for the 78  discs in the sample of UpperSco where emission from $^{12}$CO $J=3-2$ was detected at $>5\sigma$ level. \zagaria \, used the method introduced in \cite{trapman2025_size}, where the gas disc size is measured from unconvolved best fit models  to the image. From the 78 discs, $R_{\rm{CO}}$ could be measured in  67 discs and 11 of them only have upper limits. The top panel of Fig.~\ref{fig:RCO-Rcont} shows the $R_{\rm{CO}}$ radius obtained from the CO analysis vs. the dust disc size ($R_{90}$) for the discs for which, both the dust continuum and the CO are detected.  

The median value of the $R_{\rm{CO}}$/$R_{90}$ ratio for all the resolved discs is $2.7^{+0.3}_{-0.4}$. If the calculation is split between structured and smooth discs, the values of $R_{\rm{CO}}$/$R_{90}$  are $2.3^{+0.4}_{-0.5}$ and $4.1^{+0.5}_{-1.8}$, respectively (errors correspond to the 68\%  confidence interval). To check if the distribution of the  $R_{\rm{CO}}$/$R_{90}$ ratio is significantly different in the structured vs. the smooth discs, we compare the two distributions and calculate the $p-$value from the \texttt{logrank} test from \texttt{lifelines}, finding that is much lower than 1\%, meaning that these two distributions are significantly different in our UpperSco sample  (bottom panel of Fig.~\ref{fig:RCO-Rcont}).

The $R_{\rm{CO}}$/$R_{90}$ ratio is in all cases higher than unity and in 10 discs is higher than 4. Ratios of $R_{\rm{CO}}$/$R_{90}$ lower than 4 can explained by the difference in optical depth between the optically thick $^{12}$CO  and the more optically thin continuum  emission \citep{dutrey1998, Guilloteau1998, fachhini2017}. The disc with the highest $R_{\rm{CO}}$/$R_{90}$ is 2MASS-J16072955-2308221 with a value of $10.8^{+2.9}_{-3.2}$, and the disc with the lowest $R_{\rm{CO}}$/$R_{90}$ is  2MASS-J16140792-1938292 with a value of $1.1^{+0.04}_{-0.004}$. These two discs are not particularly different from the rest of the discs in terms of CO flux. Of the 10 that discs have values $R_{\rm{CO}}$/$R_{90}>4$,  all of them are classified as smooth in our sample, which is consistent with radial drift of dust particles taking place and making the dust disc much smaller than the gaseous disc \citep{trapman2019, kurtovic2021}. Nonetheless, the interpretation of $R_{\rm{CO}}$/$R_{90}$ in dust evolution models with and without pressure traps is complex, as it is shown in Sect.~\ref{sect:dust_evolution_results}. 

\begin{figure*}
    \centering
    \includegraphics[width=\linewidth]{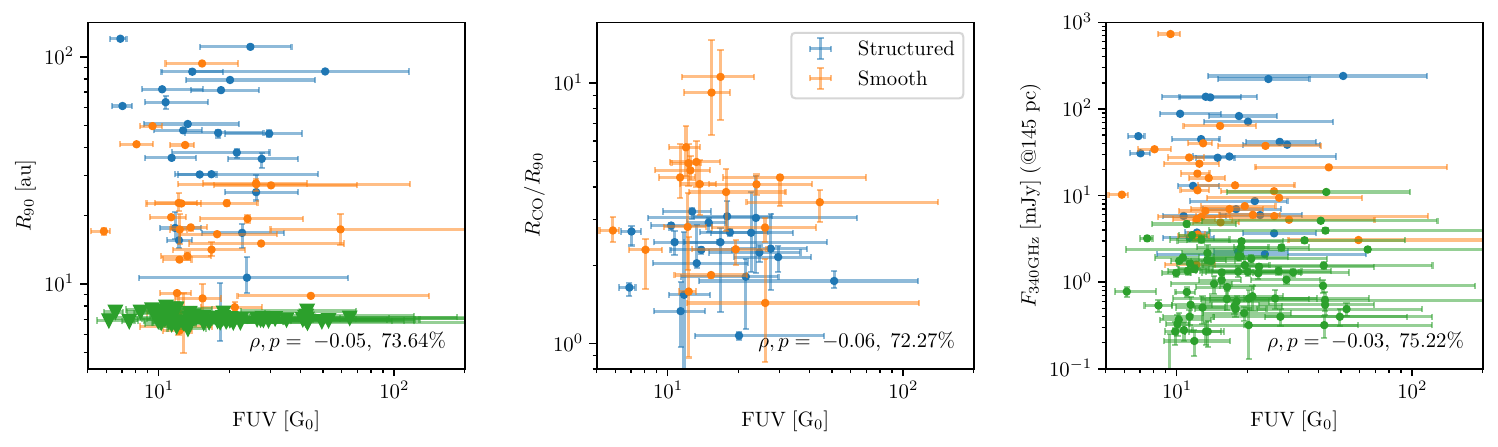}
    \caption{Disc properties against $F_{\rm{UV}}$. Left: Dust continuum radius. Middle: Gas to dust size ratio. Right: Dust continuum flux at Band 7. Blue, orange, and green points represent structured discs, smooth discs, and unresolved discs, respectively.}
    \label{fig:FUV_corr}
\end{figure*}

\cite{long2022} analysed $R_{\rm{CO}}$/$R_{90}$  of a sample of 44 discs in several nearby star-forming regions, mostly Taurus, Lupus, and Ophiucus and found an averaged value of $R_{\rm{CO}}$/$R_{90} =2.9\pm1.2$. We compare the CDFs of the sample in \cite{long2022} and our sources, and we did not find a significant variation of the distribution of the $R_{\rm{CO}}$/$R_{90}$  ratio between younger star forming regions and the value we find in UpperSco. In addition, \cite{long2022} did not find a significant difference of $R_{\rm{CO}}$/$R_{90}$ between discs with structures and smooth discs. Similarly, in \cite{trapman2025_size} there are no significant difference in $R_{\rm{CO}}$/$R_{90}$ between Lupus and UpperSco, and no difference between structured and smooth discs. The difference of $R_{\rm{CO}}$/$R_{90}$  between smooth and structured discs is significant in our results, contrary to \cite{long2022} and \cite{trapman2025_size}.

We look at the distribution of $R_{90}$, and $R_{\rm{CO}}$/$R_{90}$ for the structured and smooth discs based on the spectral type of the central star (Fig~\ref{fig:R90_SpT}, top panels). There is a small trend of $R_{90}$ decreasing for late type stars in both groups (structured and smooth discs), which leads to a modest trend of $R_{\rm{CO}}$/$R_{90}$ increasing for late type stars. In addition, we perform a similar analysis of the distribution of discs per the sub-clusters in UpperScorpius OB Association  \citep[described in][]{Ratzenbock2023a}, to check if the distribution of dust disc sizes changes across clusters, and therefore different ages between $\sim4$ to $\sim14$\,Myr. We do not find any trend for $R_{90}$ with age, but for $R_{\rm{CO}}$/$R_{90}$  there is a slight decrease for the smooth discs, while $R_{\rm{CO}}$/$R_{90}$ is approximately constant for structured discs,  as shown in the bottom panels of Fig.~\ref{fig:R90_SpT}.

Finally, we look for potential correlations of $R_{90}$ and $R_{\rm{CO}}$/$R_{90}$, as well as the millimetre fluxes,  with environmental UV flux ($G_0$). The $F_{\rm{UV}}$ values are evaluated as the sum of the contribution of the surrounding OBA-type stars, where we accounted for the uncertainty in the separation between disc and massive stars in a 3D space. Specifically,  given the 2D geometry of the stellar cluster, the local density function of the region can be defined, and therefore the probability distribution of the 3D separation from massive stars. We sampled from this probability to compute the distribution of the FUV flux \citep{anania2025b}. We do not find any potential correlation of these quantities with $F_{\rm{UV}}$ (Fig.~\ref{fig:FUV_corr}) in agreement with the results from \zagaria~for $R_{\rm{CO}}$.

%%%%%%%%%%
\subsection{Disc dust mass vs. stellar mass relationship}
%%%%%%%%%%
We  derived the stellar mass for all possible targets in this work. For this, the effective temperatures are obtained from spectral types from \cite{manara2020} if available, otherwise from \cite{luhman2022} using the conversions of \cite{Pecaut2013}. After correcting for extinction \citep[using the extinction law from] [and 2MASS and DENIS photometry]{cardelli1989}, stellar luminosities are derived by scaling the 2MASS J observed photometries to the Gaia DR3 distances \citep[using the geometric distances of][]{bailer20221}. To derive stellar masses, the stellar luminosities and effective temperatures are then compared to an interpolated grid of \cite{baraffe2015} pre-main sequence evolutionary tracks. A conservative 10\% error in stellar luminosity and effective temperature is applied to all sources. For 94 of the 285 sources of the whole sample in \cite{carpenter2025}, no luminosity could be obtained and hence no stellar mass is estimated because either there is a lack of data to compute effective temperatures or stellar luminosities, or because they fall outside pre-main sequence tracks in the HR diagram (the latter is only the case for two sources).

\begin{figure}
    \centering
    \includegraphics[width=1.0\linewidth]{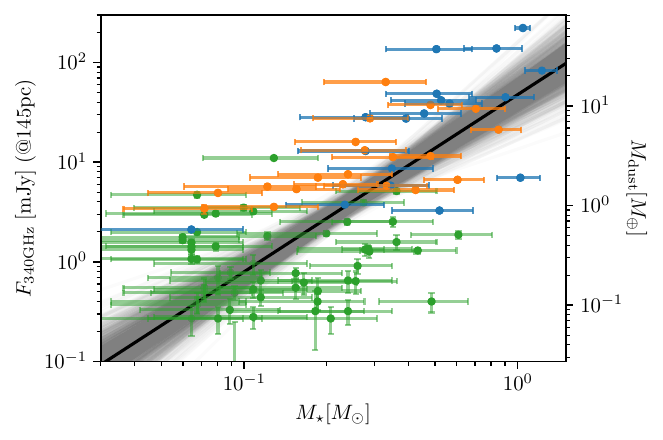}
    \caption{Flux (left y-axis) or $M_{\rm{dust}}$ (right y-axis) vs. the stellar mass. The fluxes are normalised to a distance of 145,\,pc. Blue, orange, and black points represent the structured, smooth discs, and unresolved discs, respectively.}
    \label{fig:Mdust_Mstar}
\end{figure}

In terms of number of discs for which we can calculate stellar masses and for which we perform visibility analysis, we have the following:  18 of the 24 structured discs, 20 of the 28 smooth discs,  and 54 of the 69  unresolved discs, for a total overlap of 92 discs in our sample. Figure~\ref{fig:Mdust_Mstar} shows the correlation between the flux at 340\,GHz (left y-axis) or $M_{\rm{dust}}$ (right y-axis) vs. the stellar mass. For the disc mass, we assume optically thin emission \citep{HILDEBRAND1983}, that is:

\begin{equation}
        M_{\mathrm{dust}}=\frac{{d^2 F_\nu}}{\kappa_\nu B_\nu (T)}.
  \label{mm_dust_mass}
\end{equation}

\noindent For $\kappa_\nu$, we assume  $\kappa_\nu=2.3\,$cm$^{2}$\,g$^{-1}(\nu/230\,\rm{GHz})^{0.4}$ \citep[e.g.,][]{andrews2013}, and $B_\nu$ is the blackbody surface brightness assuming a disc temperature of 20\,K. 

\begin{figure}
    \centering
\includegraphics[width=0.9\linewidth]{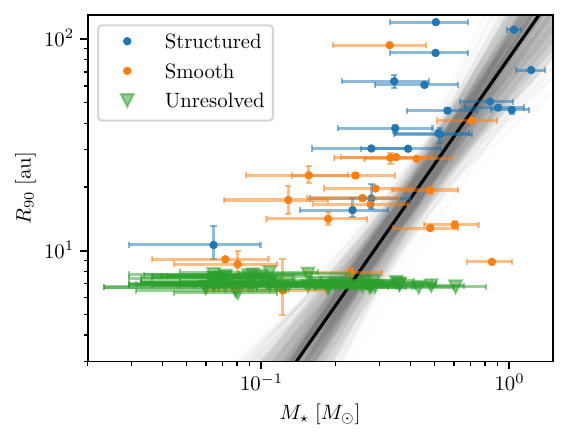}
    \caption{Dust disc size $R_{90}$ vs. stellar mass. Blue, orange, and black points represent the structured, smooth discs, and unresolved discs, respectively.}
    \label{fig:R90_Mstar}
\end{figure}

\begin{figure*}
\centering
\includegraphics[width=\linewidth]{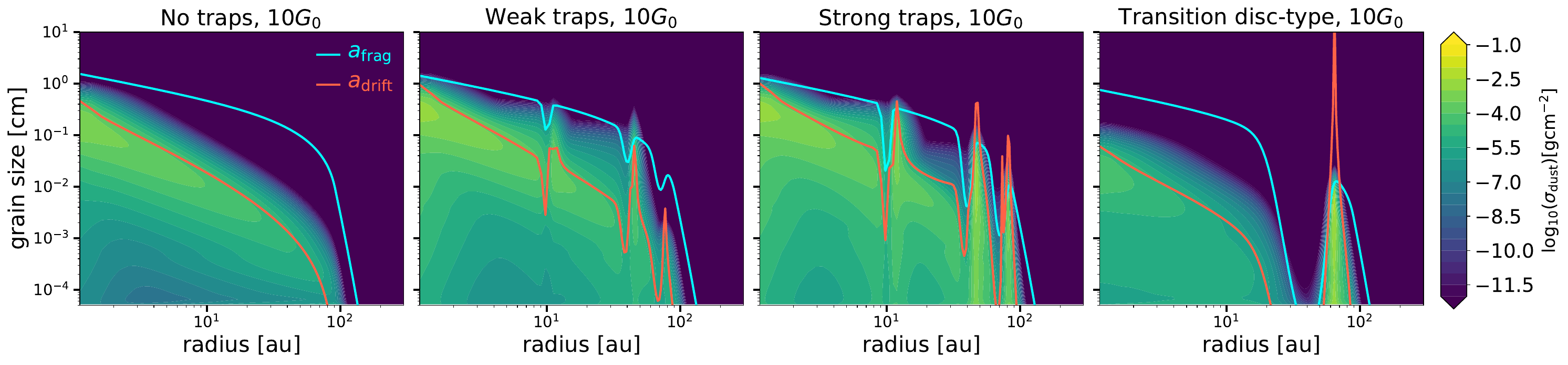}
\caption{Dust density distribution after 2\,Myr of evolution for discs with (from left to right): no traps, weak traps, strong-traps, and transition disc-type trap. All of these models assume a $F_{\rm{UV}}=10\,G_0$. These models are for the case of discs around a 0.3\,$M_\odot$ star.}
\label{fig:dust_density}
\end{figure*}

Previous works reported the $M_{\rm{dust}}-M_\star$ for several star forming regions \citep[e.g.,][]{andrews2013, pascucci2016, ansdell2017}, finding that this correlation is similar to the young star-forming regions of  Taurus, Lupus, and Chamaeleon I ($M_{\rm{dust}}\propto M_\star^{\sim1.2-1.6}$), that are $\sim$1-3\,Myr, but that steepens for the older star-forming region  UpperSco ($M_{\rm{dust}} \propto M_\star^{\sim 2.4-2.7}$).  We fit the $M_{\rm{dust}}-M_\star$ correlation using our data and including the unresolved discs, and the results are summarised in Table~\ref{Table:summary_fits}. Our best fit correlations is less steep than reported previously for UpperSco \citep{baranfeld2016, pascucci2016}, and similar to the values found in younger regions.

As the dust continuum flux is known to scale with stellar mass (in this work as $F_{\rm{340GHz}}\propto M_\star^{1.83} $), and the disc millimetre flux scales with dust disc radius (in this work as $R_{90}\propto F_{\rm{340GHz}}^{0.72}$, Sect.~\ref{sect:Lmm-R90}), it is expected that there is a correlation with $M_\star$ and $R_{90}$. Figure~\ref{fig:R90_Mstar} shows the correlation between the dust disc size $R_{90}$ and stellar mass ($R_{90}\propto M_\star^{1.69}$), and the results of the fit are in Table~\ref{Table:summary_fits}. From the previous relations, one expects $R_{90} \propto M_*^{1.83*0.72}$, which is consistent within the uncertainties with the slope reported in Table~\ref{Table:summary_fits}. This correlation is strong and with very low chance to be random, as suggested by the $\rho$ values. This correlation was not found in most of the star forming regions analysed by \cite{hendler2020}, but reported in \cite{andrews2018} for Lupus and Taurus using Band 7 ALMA observations, with a much flatter slope of $\sim0.6$. As explained in \cite{andrews2018}, depending on if the dust emission is optically thick or thin, the expected correlation between $M_\star$ and $R_{90}$ can be recovered with different slopes, although it is not straightforward in the case of optically thin emission as it would depend on the scaling relations of the disc temperature, dust surface density and optical depth. As the slope is higher in UpperSco compared  to the values reported for younger discs in Lupus and Taurus, we suggest that this is because discs in UpperSco are less optically thick. Finally, we do not identify a clear difference between smooth and structured discs in these correlations of $M_{\rm{dust}}-M_\star$ or $R_{90}-M_{\star}$.

\subsection{Gas and Dust Evolution Models} \label{sect:dust_evolution_results}

Figure~\ref{fig:dust_density} shows an example of the dust density distribution  as a function of grain size and distance from the star from the dust evolution simulations with $F_{\rm{UV}}=10\,G_0$. These are shown for discs that are considered smooth (no traps) and discs with structures (weak-, strong-, and transition disc-type traps). The models shown in Fig.~\ref{fig:dust_density} correspond to discs around a 0.3\,$M_\odot$ stellar mass. The results for 1.0\,$M_\odot$ are shown in Fig~\ref{fig:dust_density_1Msun}.  All the models are shown at the same time of evolution, which is 2\,Myr, which means 1\,Myr after external photoevaporation has been active. This figure shows the effect of dust trapping for a given $G_0=10$, where traps help to accumulate dust in pressure maxima, keeping dust particles in the outer disc.

\begin{figure*}
\centering
\includegraphics[width=\linewidth]{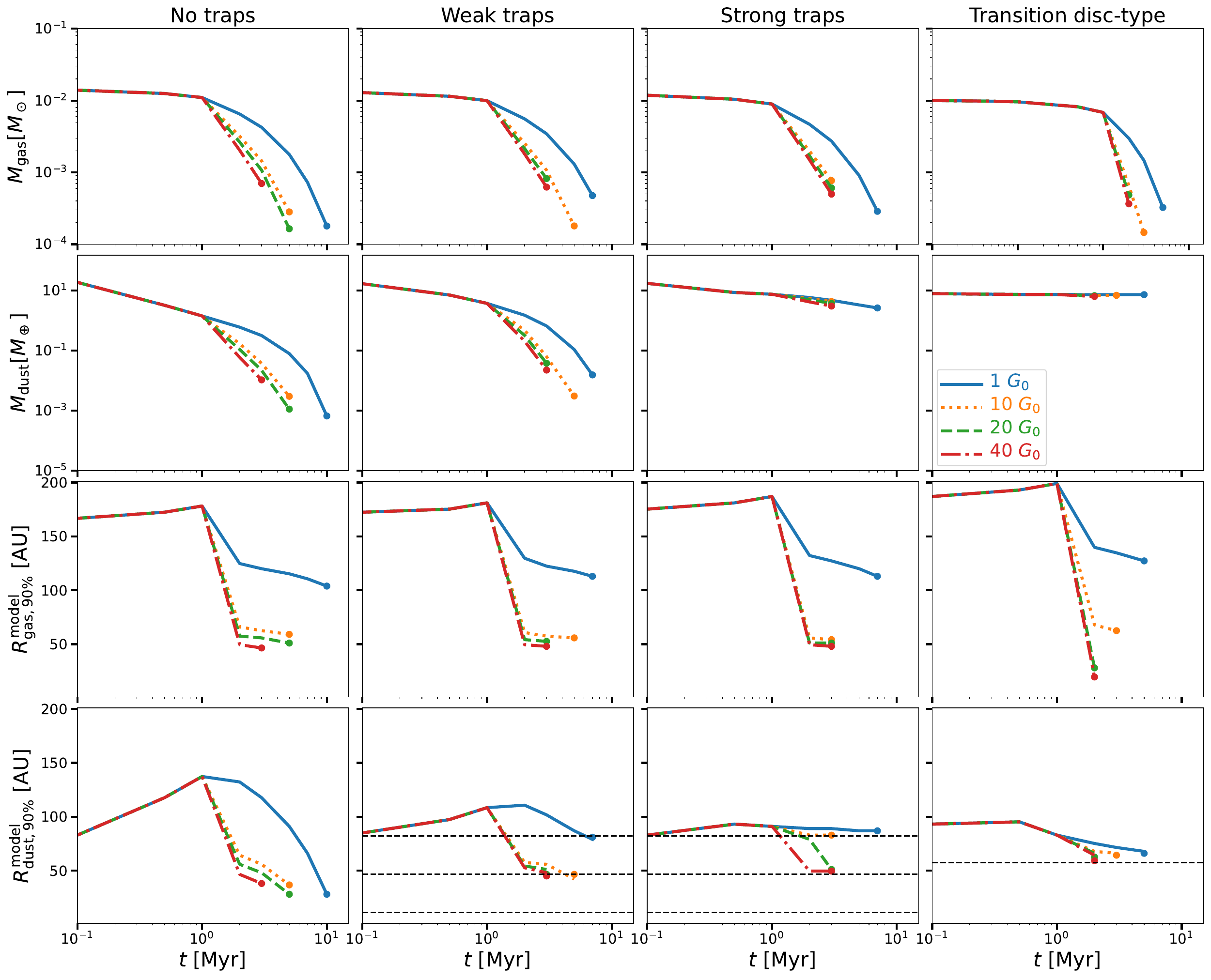}
\caption{From top to bottom: evolution of the gas disc mass, dust disc mass, gas disc size, and dust disc size directly obtained from models with different values of $F_{\rm{UV}}$. From left to right: no traps, weak traps, strong-traps, and transition-disc type trap. These are the results for the cases of a disc around a 0.3\,$M_{\odot}$. The dots represent the last time step before the disc dissipates. Horizontal dashed lines are the location of the pressure maxima in the models with dust traps.}
\label{fig:properties_models_03Msun}
\end{figure*}

\begin{figure*}
\centering
\includegraphics[width=\linewidth]{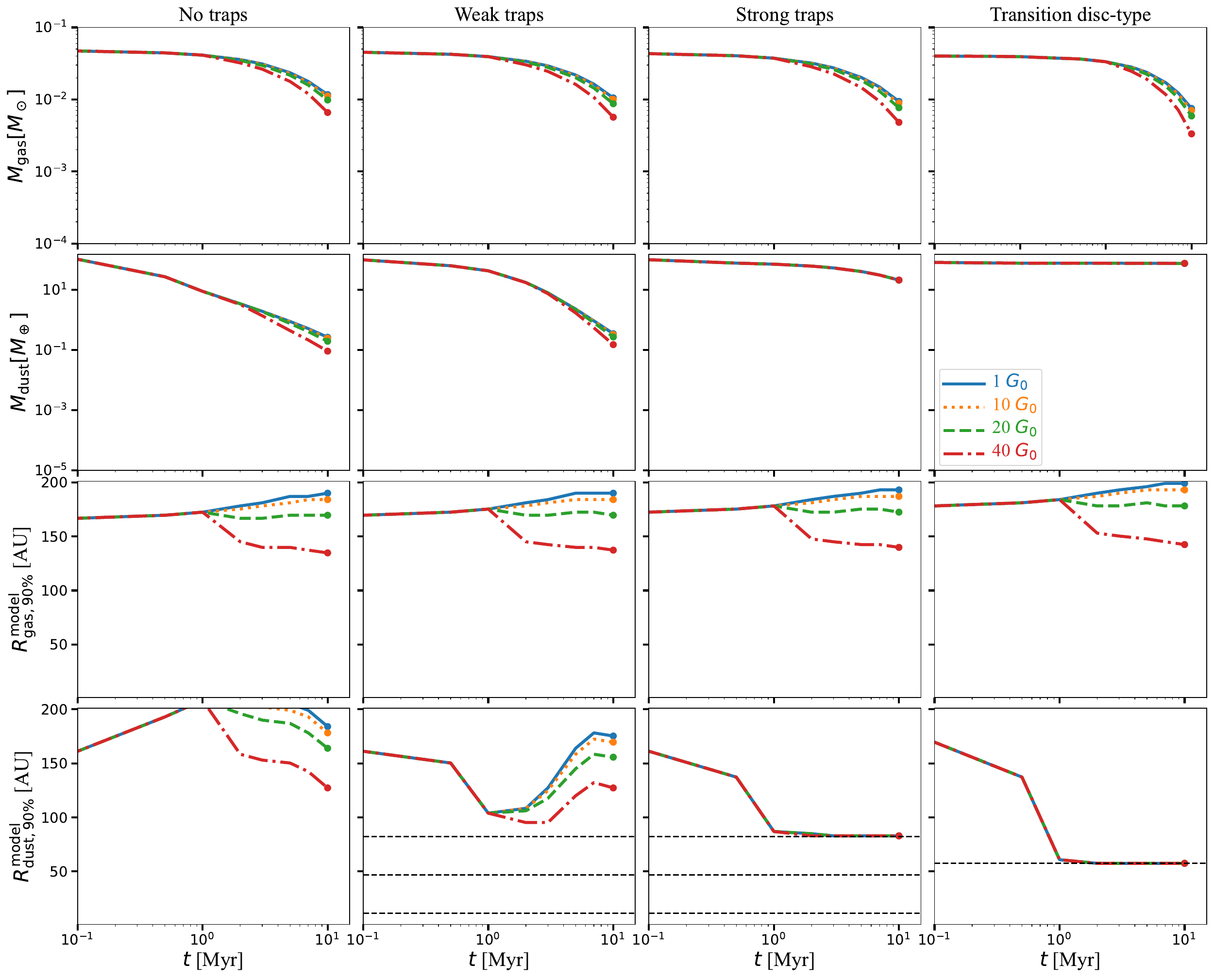}
\caption{Same as Fig~\ref{fig:properties_models_03Msun} but for a 1\,$M_\odot$ star.}
\label{fig:properties_models_1Msun}
\end{figure*}

To investigate the potential effect of $F_{\rm{UV}}$ in the models, we investigate in Figs.~\ref{fig:properties_models_03Msun} (for the case of discs around 0.3\,$M_\odot$) and ~\ref{fig:properties_models_1Msun} (for the case of discs around 1.0\,$M_\odot$)  the evolution of the gas and dust disc mass directly obtained from the models, and the evolution of the gas and dust disc size as the location that encloses 90\% of the total gas and dust mass in the models. For the dust disc sizes, we assume the contribution of all the grain sizes in the simulations. In some cases, in the models of discs around a 0.3\,$M_\odot$ star, the disc disperses before the end of the simulation. We assume  disc dispersion time when the disc gas mass is lower than 1\% of the initial gas disc mass. The time-step when this occurs is marked as a point in the figures. In the case of discs around a 0.3\,$M_\odot$ star, discs in 9/16 simulations dispersed before 5\,Myr, while in the case of discs around a 1.0\,$M_\odot$ star, all discs still exist by the last time step of the simulations (10\,Myr), in which case we also mark this last time step as point in the figures. 

Figure~\ref{fig:properties_models_03Msun} shows that once external photoevaporation is active, the disc gas mass and size decreases sharper with time for models of discs around a 0.3\,$M_\odot$ star than in discs around a 1\,$M_\odot$ star (Fig.~\ref{fig:properties_models_1Msun}). This is a natural consequence of external photoevaporation as disc material is less gravitationally bound around a lower mass star, which makes it easier to disperse the disc by photoevaporation from the UV irradiation in the environment. Interestingly, for the models of discs around a 0.3\,$M_\odot$ star,  there is a large difference of $M_{\rm{gas}}$ between $F_{\rm{UV}}=1\,G_0$ and the rest of the $F_{\rm{UV}}$ values ($F_{\rm{UV}}=10, 20, 40\,G_0$). But results do not change significantly among the models where $F_{\rm{UV}}$ is higher than 1\,$G_0$, and already for $F_{\rm{UV}}=10\,G_0$ the disc disperses by the ages of UpperSco. The dispersal time changes among models with different type of traps. However, this is a consequence of the method used to impose the traps in the disc evolution, which is by varying locally the $\alpha$-viscosity. This impacts the gas accretion rate and hence the disc lifetime. In addition, as the time steps saved in the simulations are largely separated, it is difficult to recover the exact dispersion time, and therefore the difference in dispersion timescales among models assuming different pressure traps should be taken with caution. 

The evolution of the gas disc size also shows a fast decrease once the disc is truncated by the external photoevaporation. The truncation radius corresponds to the location where the wind transitions from optically thin to optically thick at UV wavelengths. In the simulations of discs around a 0.3\,$M_\odot$ star, this truncation happens right after external photoevaporation is active, which means right after 1\,Myr of evolution, independent of the type of the value of $G_0$ and the type of pressure traps. \cite{anania2025_age-pro} demonstrated that with these low values of $G_0$ the timescale when disc truncation happens is not influenced by when external photoevaporation starts to influence the disc evolution. For testing this, \cite{anania2025_age-pro} performed simulations in which the external $F_{UV}$ is turned-on after 1\, Myr (as in our models) and in this case the gas radius rapidly evolves to the same value when the disc experiences the same external radiation from earlier times of evolution, thus the same values are obtained over million-year time scales. 

\begin{figure*}
\centering
\includegraphics[width=\linewidth]{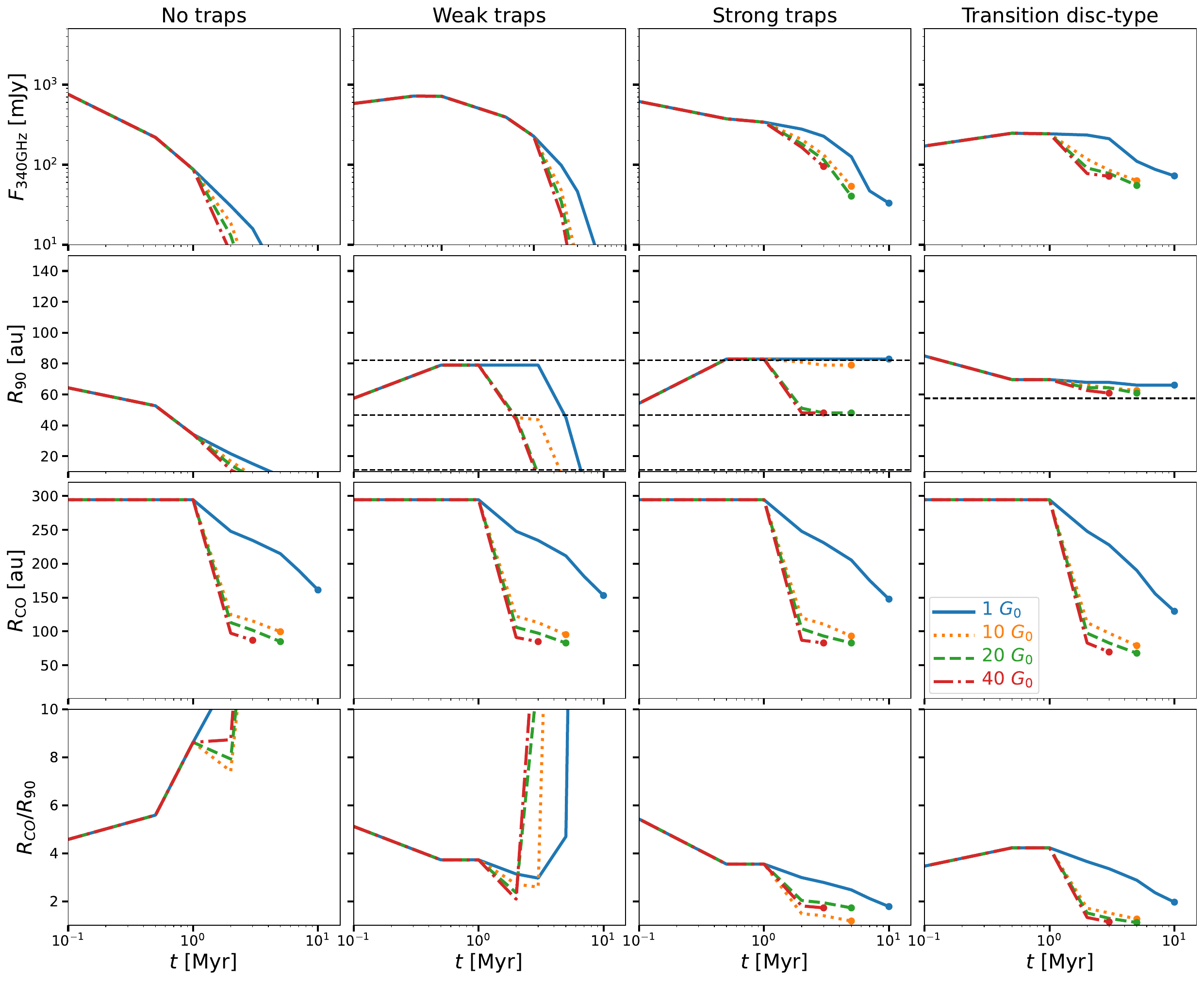}
\caption{From top to bottom: evolution of the millimetre flux, $R_{90}$, $R_{\rm{CO}}$ and $R_{\rm{CO}}/R_{90}$ for different values of $F_{\rm{UV}}$. From left to right: no traps, weak traps, strong-traps, and transition-disc type trap. Horizontal dashed lines are the location of the pressure maxima in the models with dust traps. These results correspond to the models of discs around a 0.3\,$M_\odot$ star.}
\label{properties_observables_03Msun}
\end{figure*}

Comparing the evolution of the gas mass with the case of discs around a 1.0\,$M_\odot$ star, we see that the effect of including photoevaporation is much weaker when $M_\star=1.0\,M_\odot$ independent of the $G_0$ value. For the disc size evolution, there is a less noticeable influence of external photoevaporation, where only in the case of $F_{\rm{UV}}=40\,G_0$ the gas disc size decreases slowly with time after 1\,Myr. For the other $F_{\rm{UV}}$ values, the gas disc either continues expanding or the gas size remains approximately the same until the end of the simulations.  This difference in the evolution of the gas disc size between 1.0\,$M_\odot$ and 0.3\,$M_\odot$ is mainly due to less gravitationally bound material around $M_\star=0.3\,M_\odot$. The cut-off radius in both simulations is assumed the same, but as shown by \cite{garate2024} and \cite{anania2025_age-pro}, it does not have a strong effect in the gas evolution. 

For the evolution of the dust mass and size, dust mass is more depleted in the cases of $M_\star=0.3\,M_\odot$ than for $M_\star=1.0\,M_\odot$. This is because radial drift is more efficient around lower mass stars \citep[Fig. 2 in][]{vanderMarel2023}. Comparing Figs~\ref{fig:properties_models_03Msun} and ~\ref{fig:properties_models_1Msun}, it is possible to see how weak and strong traps ($A=1,4$) are more inefficient in retaining dust in discs around $M_\star=0.3\,M_\odot$ \citep[see also][]{pinilla2020}. For the dust disc size, we see the effect of grain growth in the evolution, which makes the dust disc size to increase up to around $~$1\,Myr, and in the absence of traps to quickly shrink with time once grains have reached sizes that effectively drift. This decrease of the disc size is less sharp with time when the pressure bumps are of higher amplitude, retaining dust more efficiently. In the case of  $M_\star=0.3\,M_\odot$, the dust disc size (directly obtained in the models assuming all grains sizes) continues to decrease in all the different types of traps that are assumed, converging to the location of one of the pressure maxima, depending on the values of $F_{\rm{UV}}$. For $F_{\rm{UV}}=1, 10\,G_0$ the dust disc size converges to the location of the furthest pressure bump, while for higher $F_{\rm{UV}}$, it converges to the middle pressure bump. This is discussed in more details in  Sect.~\ref{sect:synthetic_obs}.

Contrary, in the case of $M_\star=1.0\,M_\odot$, the dust disc size decreases in the presence of weak traps, but increases again with time. This is because we are assuming all grain sizes from the models in the calculation of the dust disc size, and the micron-sized particles  contribute more to  the total disc mass  over time in the outer regions, creating an apparent increase of the dust disc size. For strong and transition-disc type traps, the dust disc size converges to nearly the location of the outer trap (or in the case of the transition disc, to the only pressure maxima present in the disc). In general, there is a low influence of $G_0$ for the evolution of the dust disc size and mass for the $M_\star=1.0\,M_\odot$ case.

\subsection{Synthetic observable disc properties} \label{sect:synthetic_obs}

Figure~\ref{properties_observables_03Msun} show the evolution of the millimeter flux at 340\,GHz, $R_{90}$, $R_{\rm{CO}}$ and $R_{\rm{CO}}/R_{90}$ for different values of $G_0$, which were calculated as described in Sect.~\ref{sect:synthetic_obs_calculations}. This figure shows the cases of  discs around a 0.3\,$M_\odot$ star with no traps, weak traps, strong-traps, and transition-disc type trap. The millimetre flux at 340\,GHz, $R_{90}$, $R_{\rm{CO}}$ evolve similar to the dust disc mass ($M_{\rm{dust}}$), the dust disc size from the models ($R_{\rm{dust}, 90\%}^{\rm{model}}$), and the gas disc size from the models ($R_{\rm{gas}, 90\%}^{\rm{model}}$), respectively (Fig.~\ref{fig:properties_models_03Msun} and \ref{fig:properties_models_1Msun}).  

For $M_\star=0.3\,M_\odot$, the millimetre continuum fluxes sharply decrease over time in the absence of traps or with weak traps, while with strong traps or transition-disc traps there is less decrease over time because particles are efficiently trapped over million-year timescales, which helps to maintain detectable levels of millimetre flux. The value of $R_{90}$ initially increases with time (within the first 1\,Myr) when millimetre-sized particles are growing in the outer disc, but once drift becomes efficient ($>$1\,Myr), $R_{90}$  decreases with time in the case of no traps or weak traps. In the case of strong traps, for the cases of $F_{\rm{UV}}=1,10\,G_0$, $R_{90}$ converges to the location of the furthest pressure bump, which is around 82\,au. However, for $F_{\rm{UV}}=20, 40\,G_0$, $R_{90}$ converges to the location of the second pressure bump around 47\,au. This is because the further out pressure bump is outside the truncation radius, which is around 50\,au  in these cases (Fig.~\ref{fig:properties_models_03Msun}). \cite{garate2024} showed that when a trap with a similar amplitude than assumed for the strong traps ($A=4$) is outside the truncation radius, it quickly dispersed for  much higher values of $G_0$ than what we assume in this work. Hence, we now demonstrate that already for $G_0\geq20$, pressure bumps with an amplitude of $A=4$ outside the truncation radius are inefficient on trapping dust particles for discs around a $0.3\,M_\odot$ star. Interestingly, for the case of a transition disc-type trap, which is of very high amplitude, the $R_{90}$ converges to the location of the only pressure trap, which is around 58\,au for all the values of $F_{\rm{UV}}$. This implies that such strong pressure makes $R_{90}$ to trace the location of the dust trap before the disc disperses, independent of the  $F_{\rm{UV}}$ value assumed for the discs in UpperSco.

The ratio of the gas to dust disc size $R_{\rm{CO}}/R_{90}$ shows very different evolution that strongly depends on the type of traps, but only slightly depends on the values of $F_{\rm{UV}}$. For no traps or weak traps, $R_{\rm{CO}}/R_{90}$ initially has values of around 4, and it continues  increasing over time, which is a result of efficient drift, and it agrees with previous results \cite[e.g.,][]{trapman2019, toci2021}. For strong traps or transition disc-type traps, the  $R_{\rm{CO}}/R_{90}$ starts with similar values of 4, but starts to decrease once millimetre-sized particles have grown and drift towards pressure maxima, while the gas disc size keeps decreasing due to the truncation by external photoevaporation. The combination of these two effects, makes  $R_{\rm{CO}}/R_{90}$ to reach values as low as $\sim1.2$ before the disc disperses.  The exact value of $R_{\rm{CO}}/R_{90}$ would depend on where pressure bumps are assumed, so the  trends when considering different traps and different $F_{\rm{UV}}$ over million-year times of evolution is the key when comparing with observations  of discs in different evolutionary stage (rather than the values themselves). 

Figure~\ref{properties_observables_1Msun} in the Appendix~\ref{app:comparison} shows the evolution of the millimetre flux at 340\,GHz, $R_{90}$, $R_{\rm{CO}}$ and $R_{\rm{CO}}/R_{90}$ for the case of $M_\star=1\,M_\odot$. The main differences with the case of $M_\star=0.3\,M_\odot$ is that because photoevaporation and drift are less effective, $R_{\rm{CO}}$ decreases less with time and $R_{90}$ converges to the location of the further dust trap independent of the amplitude of the bumps assumed.  As a consequence, while $R_{\rm{CO}}/R_{90}$ sharply increases over time for no traps, $R_{\rm{CO}}/R_{90}$ slightly increases for the other trap cases, reaching values between $~3-5$ and remaining almost constant between 1-10\,Myr. 

\begin{figure*}
\centering
\begin{tabular}{cc} 
     \includegraphics[width=8.5cm]{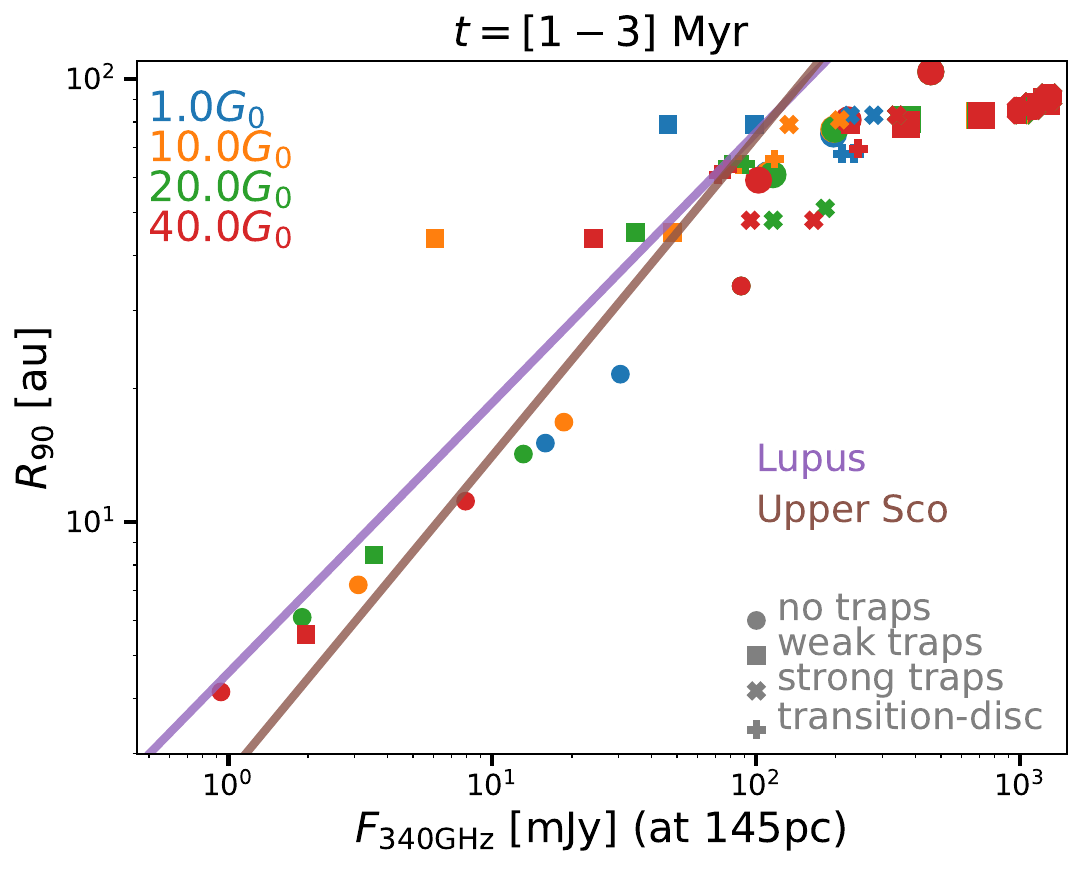}&
     \includegraphics[width=8.5cm]{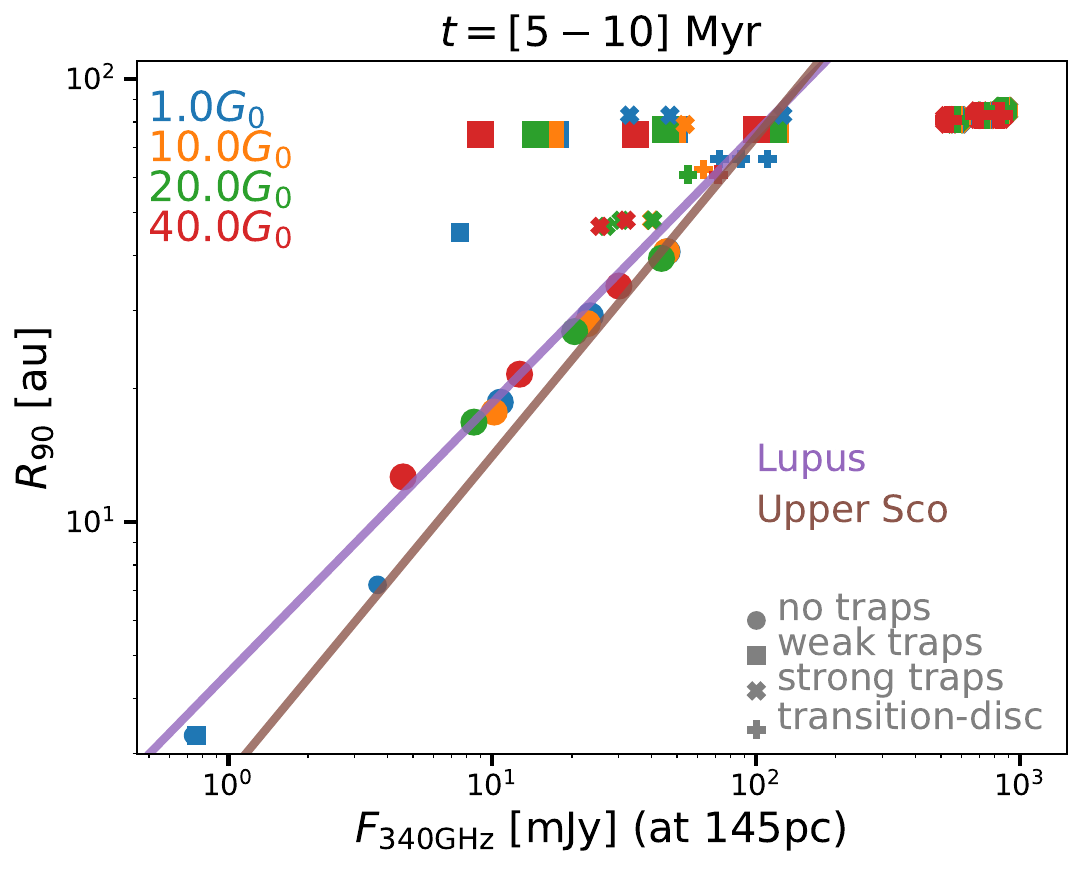}
\end{tabular}
\caption{Disc millimetre luminosity  vs. dust disc radius $R_{90}$ from the models for different values of $F_{\rm{UV}}$ and traps, assuming the outputs from 1-3\,Myr (left panel) or 5-10\,Myr (right panel). The colours represent different values of $F_{\rm{UV}}$, while the shape of the points the type of pressure traps. The size of the points symbolize different stellar masses (small size for $M_\star=0.3\,M_\odot$ and larger sizes for  $M_\star=1.0\,M_\odot$). As a reference, the best observational fit found for Lupus \citep{guerra2025} and UpperSco (this work) is overplotted.}
\label{fig:Lmm_R90_models}
\end{figure*}

Figure~\ref{fig:Lmm_R90_models} shows the disc millimetre luminosity  vs. dust disc radius $R_{90}$ from the models for different values of $F_{\rm{UV}}$, traps, $M_\star$, and different times of evolution. The main results of this figure are: (1) The values of $F_{\rm{UV}}$ do not have a strong influence on the potential correlation, as the different values (shown by different colours) are distributed everywhere in the plot, meaning that they do not drive any potential correlation in the models. (2) The presence or absence of traps has the highest impact on a potential correlation. Models with no traps or weak traps have lower millimetre fluxes and lower $R_{90}$, resembling a correlation similar to the observed ones in Lupus and UpperSco, while models with stronger traps remain with higher fluxes and dust disc sizes, potentially explaining discs located in the upper right part of the dust disc size-luminosity relationship (Fig~\ref{fig:Flux-Size}). The low mass stars drive the correlation at earlier times (1-3\,Myr), while the Solar-mass stars drive the correlation at later times (5-10\,Myr) as at this late stages the disc has dispersed in several of the models with $M_\star=0.3\,M_\odot$ as explained above.  It is important to note that although in this work we focus on investigating the effect of different values for external photoevaporation and pressure bumps, different other parameters, such as disc viscosity and fragmentation velocity can highly influence this correlation as shown in \cite{zormpas2022}.

%%%%%%%%%%%%%%%%%%%%%
\section{Discussion} \label{sect:discussion}
%%%%%%%%%%%%%%%%%%%%%

\subsection{Dust disc size, Gas-to-dust disc ratio and their evolution}
The age of the UpperSco star forming region is typically considered to be between 3 and 10\,Myr old \citep[or even older, e.g.,][]{pecaut2012, luhman2020, Armstrong2025}, which corresponds to the late stages of disc evolution \citep[e.g.,][]{fedele2010}.  
The comparison of different disc properties with younger star forming regions, such as Lupus and Taurus allow testing evolutionary effects. However, in contrast to other nearby young star forming regions, the discs in UpperSco are exposed to moderate values of $F_{\rm{UV}}$ radiation from the OBA-type stars in the association, which challenges the comparison with younger star forming regions. For this reason, it is key to compare with gas and dust evolution models that include external photoevaporation to better understand what physical mechanisms dominate the evolution and shape the observed properties.

The analysis of the morphology of the dust continuum emission of the discs resolved in UpperSco shows that discs with structures are significantly larger than smooth discs. \cite{guerra2025} reached the same conclusion when analysing a sample of discs in Lupus, which age is $\sim$1-3 Myr old. In addition, there is no significant difference of the cumulative distribution functions of the  dust disc size of the UpperSco sample when comparing with the Lupus sample from \cite{guerra2025}, suggesting no evolution of the dust disc size from the intermediate ages of discs ($\sim$1-3\,Myr) to the later stages in UpperSco ($\gtrsim$3-10\,Myr). This conclusion extends when the samples are split between structured and smooth discs (Fig~\ref{fig:CDFS}).

A more homogenous analysis of the disc dust and gas size throughout different star forming regions, including Lupus and UpperSco, was done in the AGE-PRO ALMA collaboration, which had a spatial resolution from 0.1-0.3$^{\prime \prime}$. From their analysis, they also found no significant variation of $R_{90}$ between Lupus and UpperSco \citep{vioque2025_age-pro, zhang_age-pro2025}, in agreement with the results of this work. But since the uncertainties are large because in AGE-PRO there were only 10 stars in each star forming region, there was a wide range of outcomes still possible, and their results only place an upper limit on the changes. We confirm in this study that there is no significant variation of $R_{90}$ between Lupus and UpperSco.

In contrast, the millimetre continuum fluxes are known to be lower for UpperSco compared to younger star-forming regions  \cite[see for example Fig.\,2 in][]{miotello2023}. Potential explanations of the decrease of the millimetre flux over time include dust radial drift \citep{Rosotti2019b, toci2021, stadler2022, delussu2024}, planetesimal formation \citep[e.g.,][]{Bernabo2022}, and the lost of dust particles by external factors \citep[such as photoevaporation,][]{sellek2020}. In the case of dust radial drift, it is possible that while the millimetre fluxes decrease over time, the dust disc radius remains nearly constant over time. This can happen in the presence of leaky dust traps, where the radial drift of the dust particles is reduced (not fully suppressed) since early stages ($<$1\,Myr) of the disc evolution \cite{pinilla2020} and \cite{kurtovic2025_age-pro}. If these are leaky, the observed millimetre flux decreases over time, while the dust disc size traces the location of the furthest dust trap, as also proposed by \cite{vioque2025_age-pro} to explain the lack of evolution in the observed dust disc sizes of the AGE-PRO sample, while the millimetre fluxes decrease with age. 

Our gas and dust evolution models demonstrate that these results of dust radial drift and trapping remain similar with the inclusion of mild external phototevaporation. However, this strongly depends on the mass of the hosting star, as a disc around a low mass star can be dispersed by external photoevaporation even with low values of $F_{\rm{UV}}$ irradiation ($F_{\rm{UV}}=10-40\,G_0$). In these cases, the dust traps that are located outside the photoevaporative truncation radius are destroyed, preventing the traps to keep dust particles in the outer disc and reducing the dust disc radius over time, as well as the gas disc radius. However, \cite{garate2024}  suggested that the dust particles entrained by the photoevaporative winds may shield the disc from external irradiation,  decreasing the total mass loss rate. This potential shielding is not included in our models, and it will be a subject of research in the future. 

We also investigated a potential variation of the dust disc size with the sub-regions where the discs are located \citep[tracing different ages,][bottom panels in Fig.~\ref{fig:R90_SpT}]{Ratzenbock2023b}. We do not find a significant variation of the dust disc size within the sub-regions of the UpperSco association.
This is in agreement with the models, where the discs that survive by the age of UpperSco are discs with structures, in which case no significant variation of the dust disc size is expected. However, we find a moderate decrease of $R_{\rm{CO}}$/$R_{90}$  for the smooth discs with the age of the region, contrary to model expectations;  while $R_{\rm{CO}}$/$R_{90}$ remains approximately constant for structured discs as expected from models. This highlights the importance of deeper and higher angular resolution of the gas and dust of the discs classified as smooth in this work. As for example, low contrast small structures, such as the rings observed in TW\,Hya \citep[][]{andrews2016, lee2022, das2024}, would not be resolved by the angular resolution and sensitivity of this work.

Our analysis also suggests that the millimetre fluxes, the dust disc size, and the gas-to-dust disc size ratio do not change with different values of the $F_{\rm{UV}}$ (Fig.~\ref{fig:FUV_corr}). However,  in clusters where the $F_{\rm{UV}}$ values are higher, such as Orion, $\sigma$ Orionis and NGC 2024, previous works have shown that dust disc masses decline with decreasing separation from the photoionizing source, i.e., with higher $F_{\rm{UV}}$ \citep[e.g.,][]{mann2014, mann2015, ansdell2017,  vanTerwisga2020, vanTerwisga2023}.

In our gas/dust evolution models, the gas-to-dust disc size ratio is expected to increase with time in the absence of traps. This is also the case if the traps are weak (in our models when $A=1$), and for the discs around a $0.3\,M_\odot$. Our observations suggest that the gas-to-dust disc size does not change significantly when comparing the values found in \cite{long2022} primarily for discs in Lupus and Taurus vs. the values that we found in UpperSco. The same conclusion was found in the AGE-PRO collaboration \citep{trapman2025_size, zhang_age-pro2025}. Therefore, the comparison of the models with observations suggest that traps need to be present in a significant fraction of the surviving discs.  As a $0.3\,M_\odot$ is a more representative type of star in UpperSco, this would imply that these traps need to be strong by the age of UpperSco. 

In our UpperSco sample, there are a couple of discs with extreme values of $R_{\rm{CO}}$/$R_{90}$ ($\gtrsim$6). Interestingly, two of them (2MASS-J16111705-2213085 and 2MASS-J16054540-2023088) are very low mass stars ($\sim0.08\,M_\odot$ and $\sim0.13\,M_\odot$, respectively). Our models predict that for these discs, $R_{\rm{CO}}$/$R_{90}$ would only reach those high values if no-traps or weak-traps are present (Fig.~\ref{properties_observables_03Msun}). However, they are not expected to survive more than 5\,Myr with their $F_{\rm{UV}}$ values, which are for both between 16-17\,$G_0$. A more detailed analysis of the potential initial conditions that can lead to the survival of these discs is subject in Ping et al. (in prep).

Finally, in this work, we look for any correlation between the dust disc size ($R_{90}$) and the stellar properties. We found that the dust disc size is smaller for lower mass stars, while the gas disc size does not seem to change with stellar mass, as shown is \zagaria. As a consequence, the gas-to-dust disc size ratio seems to increase for low mass stars. This could be an effect of dust drift being more efficient around lower mass stars, making the dust disc size smaller in lower mass stars \citep{pinilla2013, pinilla2022}. Interestedly, we find a correlation between $R_{90}$ and $M_\star$, which is steeper than previously reported for discs in Taurus and Lupus \citep{andrews2018}, which can be explained by less optical depth of the surviving discs in UpperSco.

\subsection{Dust disc size vs luminosity}

We find a strong relation between the disc luminosity and the dust disc size for the discs in UpperSco, which is steeper than reported in previous works for UpperSco \citep[e.g.][]{baranfeld2016, hendler2020}. When compared the correlation with the recent work from \cite{guerra2025} in Lupus, we find that this correlation is similar between Lupus and UpperSco, contrary to the results in \cite{hendler2020}.

This correlation has been analysed in different theoretical works, concluding that it can be explained by the drift of dust particles \citep[e.g.,][]{Rosotti2019b, zormpas2022}. In our models of gas/dust evolution that include external photoevaporation $F_{\rm{UV}}=1, 10, 20, 40\,G_0$ and drift (that can also be reduced or stoped by pressure traps), we demonstrated that the effect of drift dominates the correlation and that there is no significant effect of $F_{\rm{UV}}$  for the discs that survive (Fig.~\ref{fig:Lmm_R90_models}). In late stages, the correlation is driven by the discs that survive mainly around a 1.0\,$M_\odot$.  Hence, $F_{\rm{UV}}$ does have an effect on what discs populate this relationship. It is therefore important to study the survival rate of discs in UpperSco in more detail by performing population synthesis models with a more dense parameter space of disc and stellar conditions (Ping et al., in prep).

This correlation can be flattened by the presence of traps, as seen in Fig.~\ref{fig:Lmm_R90_models} and shown in \cite{pinilla2020}. Our results show that the flattening of this relation is very effective in the presence of strong traps or transition disc-type traps (Fig.~\ref{fig:Lmm_R90_models}). This sharp transition between steep (for none traps) and flat (for strong traps) may be smoother when assuming traps that become stronger over time and potential shielding due to dust particles in the wind that can help for the survival of discs around low mass stars. At the current resolution of our observations, we  find a significant difference of the slope between structured vs. smooth discs in the UpperSco sample, with the structured discs showing a flatter relation (Table~\ref{Table:summary_fits}), in agreement with model expectations. 

\section{Conclusions} \label{sect:conclusions}
In this work, we model the visibilities of the dust continuum emission of 121 discs in UpperSco from \cite{carpenter2025}, to constrain their flux, size, and geometry. Our visibility model assumes either a simple Gaussian profile with free centreing, or point sources for the unresolved and faint discs. We compare the observations with gas and dust evolution models that include external photoevaporation. Our main conclusions are:

\begin{itemize}
    \item In our sample, 20\% of the discs are structured, 23\% remain smooth, and 57\% remain unresolved with the current resolution of the observations (0.1$^{\prime \prime}$-0.3$^{\prime \prime}$). The median value of $R_{90}$ of the whole sample is $25.4^{+10.2}_{-7.7}$\,au. When compared with the recent work of \cite{guerra2025}, there is no significant difference of the cumulative distribution functions of Lupus and UpperSco, meaning that there is no a significant difference of $R_{90}$ among the discs in Lupus and the surviving discs in UpperSco. In our analysis, there is a significant difference between the CDFs of the structured vs. smooth discs in UpperSco, where the structured discs are significantly larger than the smooth discs (Table~\ref{Table:R90_R68}).

    \item The dust disc size vs. luminosity relationship in UpperSco is such that $R_{90}\propto L_{\rm{mm}}^{0.72}$, which is steeper than reported in previous works that analysed a smaller sample of UpperSco.  This correlation is similar than in Lupus, where it is $R_{90}\propto L_{\rm{mm}}^{0.61}$, suggesting that this correlation does not change with disc evolution. 

    \item In our UpperSco sample, smooth discs have a significantly larger gas to dust disc size ratio ($R_{\rm{CO}}/R_{90}$) than structured discs. There is not a significant difference of $R_{\rm{CO}}/R_{90}$ when compared to the values found in younger discs \citep[e.g.,][]{long2022}, in agreement with \cite{trapman2025_size}.   In addition, there is the modest trend of $R_{90}$ decreasing for very low mass stars in both groups (structured and smooth discs). Such a trend is not seen in $R_{\rm{CO}}$ (\zagaria), and hence a modest trend of $R_{\rm{CO}}/R_{90}$ increasing for low mass stars is driven by the changes of $R_{90}$. 
    
    \item There are no significant differences of $R_{90}$ across the different sub-clusters that span ages between $\sim 4-14\,$Myr. $R_{\rm{CO}}/R_{90}$ seems to slightly decrease with age for the smooth discs, while it remains approximately constant for structured discs.  Finally, no significant trends exist of $R_{90}$ or  $R_{\rm{CO}}/R_{90}$ vs. the environmental UV flux.

    \item We find a dust disc mass vs. stellar mass relationship, such that $M_{\rm{dust}}\propto M_\star^{1.83}$, which is flatter than previously reported for UpperSco, and it is similar to the correlation found in younger star-forming regions. We find a clear correlation between $R_{90}$ vs. stellar mass ($R_{90}\propto M_\star^{1.69}$), which is steeper than in Lupus and Taurus \citep[$R_{90}\propto M_\star^{0.6}$,][]{andrews2018}, suggesting that the surviving discs in UpperSco are less optically thick, which could be tested with future multi-wavelengths observations.

    \item The gas and dust evolution models that include mild values of $F_{\rm{UV}}$, show that discs around a low mass star ($0.3\,M_\odot$) should have already dispersed by the age of UpperSco with the initial conditions assumed in this work. A population synthesis approach is needed to quantify the dispersion rate due to external photoevaporation with a larger parameter space exploration than in this work (Ping et al. in prep). For dust disc size vs. luminosity relationship, our models demonstrate that  the values of $F_{UV}$ do not have a strong influence on this correlation, and the presence or absence of traps (that controls the level of drift) has the highest impact on this correlation. In the models, the dust disc size vs. luminosity relationship flattens when dust drift is reduced or suppressed in the presence of dust traps. This agrees with observations as this correlation for the structured discs is significantly flatter than for the smooth and unresolved discs.

    \item The evolution of $R_{\rm{CO}}/R_{90}$  from the models is that it increases with age in the absence of pressure traps. For a $1\,M_\odot$ star, the presence of weak traps is enough to keep $R_{\rm{CO}}/R_{90}$ nearly constant over the last ten million years of the disc evolution. Nonetheless, this is not the case for discs around a $0.3\,M_\odot$ star, for which only strong traps keep the $R_{\rm{CO}}/R_{90}$ value low and constant in the last million years of evolution. Comparing with observations, this suggests that for the surviving discs in UpperSco, most of the traps should be stronger than in earlier ages.

    \item Higher resolution observations are needed to confirm several of the observed trends and values, specially to compare to surveys of younger regions that have been observed at higher resolution. It is surprising that despite models predicting that discs around 0.3\,$M_\odot$  should have been dispersed by the age of 5\,Myr, they still exist. This highlights the need of population synthesis models that include external photoevaporation in parallel to higher resolution observations, specially of the gas emission, to better understand the initial conditions, evolution and fate of these discs.
    
\end{itemize}
\section*{Acknowledgments}
We are very thankful to Francesco Zagaria, Giovanni Rosotti, and Leon Trapman for providing the gas disc sizes used in this work prior to their publication, and for the scientific discussions. 
A.S. and P.P. acknowledge funding from the UK Research and Innovation (UKRI) under the UK government’s Horizon Europe funding guarantee from ERC (under grant agreement No 101076489).
N.K. acknowledge funding from the Deutsche Forschungsgemeinschaft (DFG, German Research Foundation) - 325594231, FOR 2634/2. 
Support for F.L. was provided by NASA through the NASA Hubble Fellowship grant \#HST-HF2-51512.001-A awarded by the Space Telescope Science Institute, which is operated by the Association of Universities for Research in Astronomy, Incorporated, under NASA contract NAS5-26555.  R.A. acknowledges support from the European Union (ERC Starting Grant DiscEvol, project number 101039651) and from Fondazione Cariplo, grant No. 2022-1217. Views and opinions expressed are, however, those of the author(s) only and do not necessarily reflect those of the European Union or the European Research Council. Neither the European Union nor the granting authority can be held responsible for them.

This paper makes use of the following ALMA data: ADS/JAO.
ALMA\#2011.0.00526.S, ADS/JAO.ALMA\#2013.1,00395.S,
and ADS/JAO.ALMA\#2018.1.00564.S. ALMA is a partnership of ESO (representing its member states), NSF (USA) and NINS (Japan), together with NRC (Canada), MOST and ASIAA (Taiwan), and KASI (Republic of Korea), in cooperation with the Republic of Chile. The Joint ALMA Observatory is operated by ESO, AUI/NRAO and NAOJ.

%----------------------------------------------------------
\section*{Software}
This work made use of the following software:
Astropy \citep{astropy:2013, astropy:2018},
CASA \citep{McMullin_2007}, Dustpy \citep{stammler2022}, Emcee \citep{Foreman_2013}, Galario \citep{tazarri2018}, Matplotlib \citep{Matplotlib_2007}, Numpy \citep{Numpy_2020}, Scipy \citep{SciPy_2020}, linmix \citep{kelly2007}, lifelines \citep{Davidson-Pilon2019}.

\section*{Data availability}
The data from observations and models underlying this article will be shared on request to the corresponding author. The non-calibrated ALMA data is publicly available at \href{https://almascience.nrao.edu/aq/}{https://almascience.nrao.edu/aq/} using the project codes 2011.0.00526.S, 2013.1,00395.S, and 2018.1.00564.S. 

\bibliographystyle{mnras}
\bibliography{ref} % if your bibtex file is called example.bib

\begin{thebibliography}{}
\makeatletter
\relax
\def\mn@urlcharsother{\let\do\@makeother \do\$\do\&\do\#\do\^\do\_\do\%\do\~}
\def\mn@doi{\begingroup\mn@urlcharsother \@ifnextchar [ {\mn@doi@} {\mn@doi@[]}}
\def\mn@doi@[#1]#2{\def\@tempa{#1}\ifx\@tempa\@empty \href {http://dx.doi.org/#2} {doi:#2}\else \href {http://dx.doi.org/#2} {#1}\fi \endgroup}
\def\mn@eprint#1#2{\mn@eprint@#1:#2::\@nil}
\def\mn@eprint@arXiv#1{\href {http://arxiv.org/abs/#1} {{\tt arXiv:#1}}}
\def\mn@eprint@dblp#1{\href {http://dblp.uni-trier.de/rec/bibtex/#1.xml} {dblp:#1}}
\def\mn@eprint@#1:#2:#3:#4\@nil{\def\@tempa {#1}\def\@tempb {#2}\def\@tempc {#3}\ifx \@tempc \@empty \let \@tempc \@tempb \let \@tempb \@tempa \fi \ifx \@tempb \@empty \def\@tempb {arXiv}\fi \@ifundefined {mn@eprint@\@tempb}{\@tempb:\@tempc}{\expandafter \expandafter \csname mn@eprint@\@tempb\endcsname \expandafter{\@tempc}}}

\bibitem[\protect\citeauthoryear{{Agurto-Gangas} et~al.,}{{Agurto-Gangas} et~al.}{2025}]{agurto_age-pro2025}
{Agurto-Gangas} C.,  et~al., 2025, \mn@doi [arXiv e-prints] {10.48550/arXiv.2506.10735}, \href {https://ui.adsabs.harvard.edu/abs/2025arXiv250610735A} {p. arXiv:2506.10735}

\bibitem[\protect\citeauthoryear{{Anania} et~al.,}{{Anania} et~al.}{2025a}]{anania2025_age-pro}
{Anania} R.,  et~al., 2025a, \mn@doi [arXiv e-prints] {10.48550/arXiv.2506.10743}, \href {https://ui.adsabs.harvard.edu/abs/2025arXiv250610743A} {p. arXiv:2506.10743}

\bibitem[\protect\citeauthoryear{{Anania}, {Winter}, {Rosotti}, {Vioque}, {Zari}, {Pantaleoni Gonz{\'a}lez}  \& {Testi}}{{Anania} et~al.}{2025b}]{anania2025b}
{Anania} R.,  {Winter} A.~J.,  {Rosotti} G.,  {Vioque} M.,  {Zari} E.,  {Pantaleoni Gonz{\'a}lez} M.,   {Testi} L.,  2025b, \mn@doi [\aap] {10.1051/0004-6361/202453011}, \href {https://ui.adsabs.harvard.edu/abs/2025A&A...695A..74A} {695, A74}

\bibitem[\protect\citeauthoryear{{Andrews}, {Rosenfeld}, {Kraus}  \& {Wilner}}{{Andrews} et~al.}{2013}]{andrews2013}
{Andrews} S.~M.,  {Rosenfeld} K.~A.,  {Kraus} A.~L.,   {Wilner} D.~J.,  2013, \mn@doi [\apj] {10.1088/0004-637X/771/2/129}, \href {https://ui.adsabs.harvard.edu/abs/2013ApJ...771..129A} {771, 129}

\bibitem[\protect\citeauthoryear{{Andrews} et~al.,}{{Andrews} et~al.}{2016}]{andrews2016}
{Andrews} S.~M.,  et~al., 2016, \mn@doi [\apjl] {10.3847/2041-8205/820/2/L40}, \href {https://ui.adsabs.harvard.edu/abs/2016ApJ...820L..40A} {820, L40}

\bibitem[\protect\citeauthoryear{{Andrews}, {Terrell}, {Tripathi}, {Ansdell}, {Williams}  \& {Wilner}}{{Andrews} et~al.}{2018}]{andrews2018}
{Andrews} S.~M.,  {Terrell} M.,  {Tripathi} A.,  {Ansdell} M.,  {Williams} J.~P.,   {Wilner} D.~J.,  2018, \mn@doi [\apj] {10.3847/1538-4357/aadd9f}, \href {https://ui.adsabs.harvard.edu/abs/2018ApJ...865..157A} {865, 157}

\bibitem[\protect\citeauthoryear{{Ansdell} et~al.,}{{Ansdell} et~al.}{2016}]{ansdell2016}
{Ansdell} M.,  et~al., 2016, \mn@doi [\apj] {10.3847/0004-637X/828/1/46}, \href {https://ui.adsabs.harvard.edu/abs/2016ApJ...828...46A} {828, 46}

\bibitem[\protect\citeauthoryear{{Ansdell}, {Williams}, {Manara}, {Miotello}, {Facchini}, {van der Marel}, {Testi}  \& {van Dishoeck}}{{Ansdell} et~al.}{2017}]{ansdell2017}
{Ansdell} M.,  {Williams} J.~P.,  {Manara} C.~F.,  {Miotello} A.,  {Facchini} S.,  {van der Marel} N.,  {Testi} L.,   {van Dishoeck} E.~F.,  2017, \mn@doi [\aj] {10.3847/1538-3881/aa69c0}, \href {https://ui.adsabs.harvard.edu/abs/2017AJ....153..240A} {153, 240}

\bibitem[\protect\citeauthoryear{{Armstrong}, {Tan}, {Wright}, {Jeffries}, {Kos}, {Fiorellino}, {Buder}  \& {Barrios Lopez}}{{Armstrong} et~al.}{2025}]{Armstrong2025}
{Armstrong} J.~J.,  {Tan} J.~C.,  {Wright} N.~J.,  {Jeffries} R.~D.,  {Kos} J.,  {Fiorellino} E.,  {Buder} S.,   {Barrios Lopez} D.,  2025, \mn@doi [arXiv e-prints] {10.48550/arXiv.2505.03716}, \href {https://ui.adsabs.harvard.edu/abs/2025arXiv250503716A} {p. arXiv:2505.03716}

\bibitem[\protect\citeauthoryear{{Astropy Collaboration} et~al.,}{{Astropy Collaboration} et~al.}{2013}]{astropy:2013}
{Astropy Collaboration} et~al., 2013, \mn@doi [\aap] {10.1051/0004-6361/201322068}, \href {http://adsabs.harvard.edu/abs/2013A%26A...558A..33A} {558, A33}

\bibitem[\protect\citeauthoryear{{Astropy Collaboration} et~al.,}{{Astropy Collaboration} et~al.}{2018}]{astropy:2018}
{Astropy Collaboration} et~al., 2018, \mn@doi [\aj] {10.3847/1538-3881/aabc4f}, \href {https://ui.adsabs.harvard.edu/abs/2018AJ....156..123A} {156, 123}

\bibitem[\protect\citeauthoryear{{Bae}, {Isella}, {Zhu}, {Martin}, {Okuzumi}  \& {Suriano}}{{Bae} et~al.}{2023}]{bae2023}
{Bae} J.,  {Isella} A.,  {Zhu} Z.,  {Martin} R.,  {Okuzumi} S.,   {Suriano} S.,  2023, in {Inutsuka} S.,  {Aikawa} Y.,  {Muto} T.,  {Tomida} K.,   {Tamura} M.,  eds,  Astronomical Society of the Pacific Conference Series Vol. 534, Protostars and Planets VII. p.~423 (\mn@eprint {arXiv} {2210.13314}), \mn@doi{10.48550/arXiv.2210.13314}

\bibitem[\protect\citeauthoryear{{Bailer-Jones}, {Rybizki}, {Fouesneau}, {Demleitner}  \& {Andrae}}{{Bailer-Jones} et~al.}{2021}]{bailer20221}
{Bailer-Jones} C.~A.~L.,  {Rybizki} J.,  {Fouesneau} M.,  {Demleitner} M.,   {Andrae} R.,  2021, \mn@doi [\aj] {10.3847/1538-3881/abd806}, \href {https://ui.adsabs.harvard.edu/abs/2021AJ....161..147B} {161, 147}

\bibitem[\protect\citeauthoryear{{Baraffe}, {Homeier}, {Allard}  \& {Chabrier}}{{Baraffe} et~al.}{2015}]{baraffe2015}
{Baraffe} I.,  {Homeier} D.,  {Allard} F.,   {Chabrier} G.,  2015, \mn@doi [\aap] {10.1051/0004-6361/201425481}, \href {https://ui.adsabs.harvard.edu/abs/2015A&A...577A..42B} {577, A42}

\bibitem[\protect\citeauthoryear{{Barenfeld}, {Carpenter}, {Ricci}  \& {Isella}}{{Barenfeld} et~al.}{2016}]{baranfeld2016}
{Barenfeld} S.~A.,  {Carpenter} J.~M.,  {Ricci} L.,   {Isella} A.,  2016, \mn@doi [\apj] {10.3847/0004-637X/827/2/142}, \href {https://ui.adsabs.harvard.edu/abs/2016ApJ...827..142B} {827, 142}

\bibitem[\protect\citeauthoryear{{Bernab{\`o}}, {Turrini}, {Testi}, {Marzari}  \& {Polychroni}}{{Bernab{\`o}} et~al.}{2022}]{Bernabo2022}
{Bernab{\`o}} L.~M.,  {Turrini} D.,  {Testi} L.,  {Marzari} F.,   {Polychroni} D.,  2022, \mn@doi [\apjl] {10.3847/2041-8213/ac574e}, \href {https://ui.adsabs.harvard.edu/abs/2022ApJ...927L..22B} {927, L22}

\bibitem[\protect\citeauthoryear{{Cardelli}, {Clayton}  \& {Mathis}}{{Cardelli} et~al.}{1989}]{cardelli1989}
{Cardelli} J.~A.,  {Clayton} G.~C.,   {Mathis} J.~S.,  1989, \mn@doi [\apj] {10.1086/167900}, \href {https://ui.adsabs.harvard.edu/abs/1989ApJ...345..245C} {345, 245}

\bibitem[\protect\citeauthoryear{{Carpenter}, {Esplin}, {Luhman}, {Mamajek}  \& {Andrews}}{{Carpenter} et~al.}{2025}]{carpenter2025}
{Carpenter} J.~M.,  {Esplin} T.~L.,  {Luhman} K.~L.,  {Mamajek} E.~E.,   {Andrews} S.~M.,  2025, \mn@doi [\apj] {10.3847/1538-4357/ad8ebc}, \href {https://ui.adsabs.harvard.edu/abs/2025ApJ...978..117C} {978, 117}

\bibitem[\protect\citeauthoryear{{Cieza} et~al.,}{{Cieza} et~al.}{2019}]{cieza2019}
{Cieza} L.~A.,  et~al., 2019, \mn@doi [\mnras] {10.1093/mnras/sty2653}, \href {https://ui.adsabs.harvard.edu/abs/2019MNRAS.482..698C} {482, 698}

\bibitem[\protect\citeauthoryear{{Das}, {Kurtovic}  \& {Flock}}{{Das} et~al.}{2024}]{das2024}
{Das} S.,  {Kurtovic} N.~T.,   {Flock} M.,  2024, \mn@doi [\aap] {10.1051/0004-6361/202450278}, \href {https://ui.adsabs.harvard.edu/abs/2024A&A...689A.104D} {689, A104}

\bibitem[\protect\citeauthoryear{{Davidson-Pilon} et~al.,}{{Davidson-Pilon} et~al.}{2019}]{Davidson-Pilon2019}
{Davidson-Pilon} C.,  et~al., 2019, {CamDavidsonPilon/lifelines: v0.23.4}, \mn@doi{10.5281/zenodo.3576382}

\bibitem[\protect\citeauthoryear{{Delussu}, {Birnstiel}, {Miotello}, {Pinilla}, {Rosotti}  \& {Andrews}}{{Delussu} et~al.}{2024}]{delussu2024}
{Delussu} L.,  {Birnstiel} T.,  {Miotello} A.,  {Pinilla} P.,  {Rosotti} G.,   {Andrews} S.~M.,  2024, \mn@doi [\aap] {10.1051/0004-6361/202450328}, \href {https://ui.adsabs.harvard.edu/abs/2024A&A...688A..81D} {688, A81}

\bibitem[\protect\citeauthoryear{{Dutrey}, {Guilloteau}, {Prato}, {Simon}, {Duvert}, {Schuster}  \& {Menard}}{{Dutrey} et~al.}{1998}]{dutrey1998}
{Dutrey} A.,  {Guilloteau} S.,  {Prato} L.,  {Simon} M.,  {Duvert} G.,  {Schuster} K.,   {Menard} F.,  1998, \aap, \href {https://ui.adsabs.harvard.edu/abs/1998A&A...338L..63D} {338, L63}

\bibitem[\protect\citeauthoryear{{Facchini}, {Birnstiel}, {Bruderer}  \& {van Dishoeck}}{{Facchini} et~al.}{2017}]{fachhini2017}
{Facchini} S.,  {Birnstiel} T.,  {Bruderer} S.,   {van Dishoeck} E.~F.,  2017, \mn@doi [\aap] {10.1051/0004-6361/201630329}, \href {https://ui.adsabs.harvard.edu/abs/2017A&A...605A..16F} {605, A16}

\bibitem[\protect\citeauthoryear{{Fedele}, {van den Ancker}, {Henning}, {Jayawardhana}  \& {Oliveira}}{{Fedele} et~al.}{2010}]{fedele2010}
{Fedele} D.,  {van den Ancker} M.~E.,  {Henning} T.,  {Jayawardhana} R.,   {Oliveira} J.~M.,  2010, \mn@doi [\aap] {10.1051/0004-6361/200912810}, \href {https://ui.adsabs.harvard.edu/abs/2010A&A...510A..72F} {510, A72}

\bibitem[\protect\citeauthoryear{{Foreman-Mackey}, {Hogg}, {Lang}  \& {Goodman}}{{Foreman-Mackey} et~al.}{2013}]{Foreman_2013}
{Foreman-Mackey} D.,  {Hogg} D.~W.,  {Lang} D.,   {Goodman} J.,  2013, \mn@doi [\pasp] {10.1086/670067}, \href {https://ui.adsabs.harvard.edu/abs/2013PASP..125..306F} {125, 306}

\bibitem[\protect\citeauthoryear{{G{\'a}rate}, {Pinilla}, {Haworth}  \& {Facchini}}{{G{\'a}rate} et~al.}{2024}]{garate2024}
{G{\'a}rate} M.,  {Pinilla} P.,  {Haworth} T.~J.,   {Facchini} S.,  2024, \mn@doi [\aap] {10.1051/0004-6361/202347850}, \href {https://ui.adsabs.harvard.edu/abs/2024A&A...681A..84G} {681, A84}

\bibitem[\protect\citeauthoryear{{Guerra-Alvarado}, {van der Marel}, {Williams}, {Pinilla}, {Mulders}, {Lambrechts}  \& {Sanchez}}{{Guerra-Alvarado} et~al.}{2025}]{guerra2025}
{Guerra-Alvarado} O.~M.,  {van der Marel} N.,  {Williams} J.~P.,  {Pinilla} P.,  {Mulders} G.~D.,  {Lambrechts} M.,   {Sanchez} M.,  2025, \mn@doi [\aap] {10.1051/0004-6361/202453338}, \href {https://ui.adsabs.harvard.edu/abs/2025A&A...696A.232G} {696, A232}

\bibitem[\protect\citeauthoryear{{Guilloteau} \& {Dutrey}}{{Guilloteau} \& {Dutrey}}{1998}]{Guilloteau1998}
{Guilloteau} S.,  {Dutrey} A.,  1998, \aap, \href {https://ui.adsabs.harvard.edu/abs/1998A&A...339..467G} {339, 467}

\bibitem[\protect\citeauthoryear{{Gundlach} \& {Blum}}{{Gundlach} \& {Blum}}{2015}]{Gundlach2015}
{Gundlach} B.,  {Blum} J.,  2015, \mn@doi [\apj] {10.1088/0004-637X/798/1/34}, \href {https://ui.adsabs.harvard.edu/abs/2015ApJ...798...34G} {798, 34}

\bibitem[\protect\citeauthoryear{Harris et~al.,}{Harris et~al.}{2020}]{Numpy_2020}
Harris C.~R.,  et~al., 2020, \mn@doi [Nature] {10.1038/s41586-020-2649-2}, 585, 357

\bibitem[\protect\citeauthoryear{{Haworth}, {Clarke}, {Rahman}, {Winter}  \& {Facchini}}{{Haworth} et~al.}{2018}]{haworth2018}
{Haworth} T.~J.,  {Clarke} C.~J.,  {Rahman} W.,  {Winter} A.~J.,   {Facchini} S.,  2018, \mn@doi [\mnras] {10.1093/mnras/sty2323}, \href {https://ui.adsabs.harvard.edu/abs/2018MNRAS.481..452H} {481, 452}

\bibitem[\protect\citeauthoryear{{Haworth}, {Coleman}, {Qiao}, {Sellek}  \& {Askari}}{{Haworth} et~al.}{2023}]{haworth2023}
{Haworth} T.~J.,  {Coleman} G. A.~L.,  {Qiao} L.,  {Sellek} A.~D.,   {Askari} K.,  2023, \mn@doi [\mnras] {10.1093/mnras/stad3054}, \href {https://ui.adsabs.harvard.edu/abs/2023MNRAS.526.4315H} {526, 4315}

\bibitem[\protect\citeauthoryear{{Hendler}, {Pascucci}, {Pinilla}, {Tazzari}, {Carpenter}, {Malhotra}  \& {Testi}}{{Hendler} et~al.}{2020}]{hendler2020}
{Hendler} N.,  {Pascucci} I.,  {Pinilla} P.,  {Tazzari} M.,  {Carpenter} J.,  {Malhotra} R.,   {Testi} L.,  2020, \mn@doi [\apj] {10.3847/1538-4357/ab70ba}, \href {https://ui.adsabs.harvard.edu/abs/2020ApJ...895..126H} {895, 126}

\bibitem[\protect\citeauthoryear{{Hildebrand}}{{Hildebrand}}{1983}]{HILDEBRAND1983}
{Hildebrand} R.~H.,  1983, \qjras, \href {https://ui.adsabs.harvard.edu/abs/1983QJRAS..24..267H} {24, 267}

\bibitem[\protect\citeauthoryear{Hunter}{Hunter}{2007}]{Matplotlib_2007}
Hunter J.~D.,  2007, \mn@doi [Computing in Science \& Engineering] {10.1109/MCSE.2007.55}, 9, 90

\bibitem[\protect\citeauthoryear{{Ilee}, {Walsh}, {Jennings}, {Booth}, {Rosotti}, {Teague}, {Tsukagoshi}  \& {Nomura}}{{Ilee} et~al.}{2022}]{lee2022}
{Ilee} J.~D.,  {Walsh} C.,  {Jennings} J.,  {Booth} R.~A.,  {Rosotti} G.~P.,  {Teague} R.,  {Tsukagoshi} T.,   {Nomura} H.,  2022, \mn@doi [\mnras] {10.1093/mnrasl/slac048}, \href {https://ui.adsabs.harvard.edu/abs/2022MNRAS.515L..23I} {515, L23}

\bibitem[\protect\citeauthoryear{{Kelly}}{{Kelly}}{2007}]{kelly2007}
{Kelly} B.~C.,  2007, \mn@doi [\apj] {10.1086/519947}, \href {https://ui.adsabs.harvard.edu/abs/2007ApJ...665.1489K} {665, 1489}

\bibitem[\protect\citeauthoryear{{Kenyon} \& {Hartmann}}{{Kenyon} \& {Hartmann}}{1987}]{kenyon1987}
{Kenyon} S.~J.,  {Hartmann} L.,  1987, \mn@doi [\apj] {10.1086/165866}, \href {https://ui.adsabs.harvard.edu/abs/1987ApJ...323..714K} {323, 714}

\bibitem[\protect\citeauthoryear{{Kurtovic} et~al.,}{{Kurtovic} et~al.}{2021}]{kurtovic2021}
{Kurtovic} N.~T.,  et~al., 2021, \mn@doi [\aap] {10.1051/0004-6361/202038983}, \href {https://ui.adsabs.harvard.edu/abs/2021A&A...645A.139K} {645, A139}

\bibitem[\protect\citeauthoryear{{Kurtovic} et~al.,}{{Kurtovic} et~al.}{2025}]{kurtovic2025_age-pro}
{Kurtovic} N.~T.,  et~al., 2025, \mn@doi [arXiv e-prints] {10.48550/arXiv.2506.10740}, \href {https://ui.adsabs.harvard.edu/abs/2025arXiv250610740K} {p. arXiv:2506.10740}

\bibitem[\protect\citeauthoryear{{Lissauer}, {Batalha}  \& {Borucki}}{{Lissauer} et~al.}{2023}]{lissauer2023}
{Lissauer} J.~J.,  {Batalha} N.~M.,   {Borucki} W.~J.,  2023, in {Inutsuka} S.,  {Aikawa} Y.,  {Muto} T.,  {Tomida} K.,   {Tamura} M.,  eds,  Astronomical Society of the Pacific Conference Series Vol. 534, Protostars and Planets VII. p.~839 (\mn@eprint {arXiv} {2311.04981}), \mn@doi{10.48550/arXiv.2311.04981}

\bibitem[\protect\citeauthoryear{{Long} et~al.,}{{Long} et~al.}{2019}]{long2019}
{Long} F.,  et~al., 2019, \mn@doi [\apj] {10.3847/1538-4357/ab2d2d}, \href {https://ui.adsabs.harvard.edu/abs/2019ApJ...882...49L} {882, 49}

\bibitem[\protect\citeauthoryear{{Long} et~al.,}{{Long} et~al.}{2022}]{long2022}
{Long} F.,  et~al., 2022, \mn@doi [\apj] {10.3847/1538-4357/ac634e}, \href {https://ui.adsabs.harvard.edu/abs/2022ApJ...931....6L} {931, 6}

\bibitem[\protect\citeauthoryear{{Luhman}}{{Luhman}}{2022}]{luhman2022}
{Luhman} K.~L.,  2022, \mn@doi [\aj] {10.3847/1538-3881/ac35e2}, \href {https://ui.adsabs.harvard.edu/abs/2022AJ....163...24L} {163, 24}

\bibitem[\protect\citeauthoryear{{Luhman} \& {Esplin}}{{Luhman} \& {Esplin}}{2020}]{luhman2020}
{Luhman} K.~L.,  {Esplin} T.~L.,  2020, \mn@doi [\aj] {10.3847/1538-3881/ab9599}, \href {https://ui.adsabs.harvard.edu/abs/2020AJ....160...44L} {160, 44}

\bibitem[\protect\citeauthoryear{{Lynden-Bell} \& {Pringle}}{{Lynden-Bell} \& {Pringle}}{1974}]{lyndenbell1974}
{Lynden-Bell} D.,  {Pringle} J.~E.,  1974, \mn@doi [\mnras] {10.1093/mnras/168.3.603}, \href {https://ui.adsabs.harvard.edu/abs/1974MNRAS.168..603L} {168, 603}

\bibitem[\protect\citeauthoryear{{Manara} et~al.,}{{Manara} et~al.}{2020}]{manara2020}
{Manara} C.~F.,  et~al., 2020, \mn@doi [\aap] {10.1051/0004-6361/202037949}, \href {https://ui.adsabs.harvard.edu/abs/2020A&A...639A..58M} {639, A58}

\bibitem[\protect\citeauthoryear{{Mann} et~al.,}{{Mann} et~al.}{2014}]{mann2014}
{Mann} R.~K.,  et~al., 2014, \mn@doi [\apj] {10.1088/0004-637X/784/1/82}, \href {https://ui.adsabs.harvard.edu/abs/2014ApJ...784...82M} {784, 82}

\bibitem[\protect\citeauthoryear{{Mann}, {Andrews}, {Eisner}, {Williams}, {Meyer}, {Di Francesco}, {Carpenter}  \& {Johnstone}}{{Mann} et~al.}{2015}]{mann2015}
{Mann} R.~K.,  {Andrews} S.~M.,  {Eisner} J.~A.,  {Williams} J.~P.,  {Meyer} M.~R.,  {Di Francesco} J.,  {Carpenter} J.~M.,   {Johnstone} D.,  2015, \mn@doi [\apj] {10.1088/0004-637X/802/2/77}, \href {https://ui.adsabs.harvard.edu/abs/2015ApJ...802...77M} {802, 77}

\bibitem[\protect\citeauthoryear{{Mathis}, {Rumpl}  \& {Nordsieck}}{{Mathis} et~al.}{1977}]{mathis1977}
{Mathis} J.~S.,  {Rumpl} W.,   {Nordsieck} K.~H.,  1977, \mn@doi [\apj] {10.1086/155591}, \href {https://ui.adsabs.harvard.edu/abs/1977ApJ...217..425M} {217, 425}

\bibitem[\protect\citeauthoryear{{McMullin}, {Waters}, {Schiebel}, {Young}  \& {Golap}}{{McMullin} et~al.}{2007}]{McMullin_2007}
{McMullin} J.~P.,  {Waters} B.,  {Schiebel} D.,  {Young} W.,   {Golap} K.,  2007, in {Shaw} R.~A.,  {Hill} F.,   {Bell} D.~J.,  eds,  Astronomical Society of the Pacific Conference Series Vol. 376, Astronomical Data Analysis Software and Systems XVI. p.~127

\bibitem[\protect\citeauthoryear{{Miotello}, {Kamp}, {Birnstiel}, {Cleeves}  \& {Kataoka}}{{Miotello} et~al.}{2023}]{miotello2023}
{Miotello} A.,  {Kamp} I.,  {Birnstiel} T.,  {Cleeves} L.~C.,   {Kataoka} A.,  2023, in {Inutsuka} S.,  {Aikawa} Y.,  {Muto} T.,  {Tomida} K.,   {Tamura} M.,  eds,  Astronomical Society of the Pacific Conference Series Vol. 534, Protostars and Planets VII. p.~501 (\mn@eprint {arXiv} {2203.09818}), \mn@doi{10.48550/arXiv.2203.09818}

\bibitem[\protect\citeauthoryear{{Pascucci} et~al.,}{{Pascucci} et~al.}{2016}]{pascucci2016}
{Pascucci} I.,  et~al., 2016, \mn@doi [\apj] {10.3847/0004-637X/831/2/125}, \href {https://ui.adsabs.harvard.edu/abs/2016ApJ...831..125P} {831, 125}

\bibitem[\protect\citeauthoryear{{Pecaut} \& {Mamajek}}{{Pecaut} \& {Mamajek}}{2013}]{Pecaut2013}
{Pecaut} M.~J.,  {Mamajek} E.~E.,  2013, \mn@doi [\apjs] {10.1088/0067-0049/208/1/9}, \href {https://ui.adsabs.harvard.edu/abs/2013ApJS..208....9P} {208, 9}

\bibitem[\protect\citeauthoryear{{Pecaut}, {Mamajek}  \& {Bubar}}{{Pecaut} et~al.}{2012}]{pecaut2012}
{Pecaut} M.~J.,  {Mamajek} E.~E.,   {Bubar} E.~J.,  2012, \mn@doi [\apj] {10.1088/0004-637X/746/2/154}, \href {https://ui.adsabs.harvard.edu/abs/2012ApJ...746..154P} {746, 154}

\bibitem[\protect\citeauthoryear{{Pinilla}}{{Pinilla}}{2022}]{pinilla2022}
{Pinilla} P.,  2022, \mn@doi [European Physical Journal Plus] {10.1140/epjp/s13360-022-03384-1}, \href {https://ui.adsabs.harvard.edu/abs/2022EPJP..137.1206P} {137, 1206}

\bibitem[\protect\citeauthoryear{{Pinilla}, {Birnstiel}, {Benisty}, {Ricci}, {Natta}, {Dullemond}, {Dominik}  \& {Testi}}{{Pinilla} et~al.}{2013}]{pinilla2013}
{Pinilla} P.,  {Birnstiel} T.,  {Benisty} M.,  {Ricci} L.,  {Natta} A.,  {Dullemond} C.~P.,  {Dominik} C.,   {Testi} L.,  2013, \mn@doi [\aap] {10.1051/0004-6361/201220875}, \href {https://ui.adsabs.harvard.edu/abs/2013A&A...554A..95P} {554, A95}

\bibitem[\protect\citeauthoryear{{Pinilla}, {Pascucci}  \& {Marino}}{{Pinilla} et~al.}{2020}]{pinilla2020}
{Pinilla} P.,  {Pascucci} I.,   {Marino} S.,  2020, \mn@doi [\aap] {10.1051/0004-6361/201937003}, \href {https://ui.adsabs.harvard.edu/abs/2020A&A...635A.105P} {635, A105}

\bibitem[\protect\citeauthoryear{{Qiao}, {Haworth}, {Sellek}  \& {Ali}}{{Qiao} et~al.}{2022}]{qiao2022}
{Qiao} L.,  {Haworth} T.~J.,  {Sellek} A.~D.,   {Ali} A.~A.,  2022, \mn@doi [\mnras] {10.1093/mnras/stac684}, \href {https://ui.adsabs.harvard.edu/abs/2022MNRAS.512.3788Q} {512, 3788}

\bibitem[\protect\citeauthoryear{{Ratzenb{\"o}ck}, {Gro{\ss}schedl}, {M{\"o}ller}, {Alves}, {Bomze}  \& {Meingast}}{{Ratzenb{\"o}ck} et~al.}{2023a}]{Ratzenbock2023a}
{Ratzenb{\"o}ck} S.,  {Gro{\ss}schedl} J.~E.,  {M{\"o}ller} T.,  {Alves} J.,  {Bomze} I.,   {Meingast} S.,  2023a, \mn@doi [\aap] {10.1051/0004-6361/202243690}, \href {https://ui.adsabs.harvard.edu/abs/2023A&A...677A..59R} {677, A59}

\bibitem[\protect\citeauthoryear{{Ratzenb{\"o}ck} et~al.,}{{Ratzenb{\"o}ck} et~al.}{2023b}]{Ratzenbock2023b}
{Ratzenb{\"o}ck} S.,  et~al., 2023b, \mn@doi [\aap] {10.1051/0004-6361/202346901}, \href {https://ui.adsabs.harvard.edu/abs/2023A&A...678A..71R} {678, A71}

\bibitem[\protect\citeauthoryear{{Ricci}, {Testi}, {Natta}, {Neri}, {Cabrit}  \& {Herczeg}}{{Ricci} et~al.}{2010}]{ricci2010}
{Ricci} L.,  {Testi} L.,  {Natta} A.,  {Neri} R.,  {Cabrit} S.,   {Herczeg} G.~J.,  2010, \mn@doi [\aap] {10.1051/0004-6361/200913403}, \href {https://ui.adsabs.harvard.edu/abs/2010A&A...512A..15R} {512, A15}

\bibitem[\protect\citeauthoryear{{Rosotti}, {Booth}, {Tazzari}, {Clarke}, {Lodato}  \& {Testi}}{{Rosotti} et~al.}{2019}]{Rosotti2019b}
{Rosotti} G.~P.,  {Booth} R.~A.,  {Tazzari} M.,  {Clarke} C.,  {Lodato} G.,   {Testi} L.,  2019, \mn@doi [\mnras] {10.1093/mnrasl/slz064}, \href {https://ui.adsabs.harvard.edu/abs/2019MNRAS.486L..63R} {486, L63}

\bibitem[\protect\citeauthoryear{{Sellek}, {Booth}  \& {Clarke}}{{Sellek} et~al.}{2020}]{sellek2020}
{Sellek} A.~D.,  {Booth} R.~A.,   {Clarke} C.~J.,  2020, \mn@doi [\mnras] {10.1093/mnras/stz3528}, \href {https://ui.adsabs.harvard.edu/abs/2020MNRAS.492.1279S} {492, 1279}

\bibitem[\protect\citeauthoryear{{Shakura} \& {Sunyaev}}{{Shakura} \& {Sunyaev}}{1973}]{shakura1973}
{Shakura} N.~I.,  {Sunyaev} R.~A.,  1973, \aap, \href {http://adsabs.harvard.edu/abs/1973A%26A....24..337S} {24, 337}

\bibitem[\protect\citeauthoryear{{Sierra} et~al.,}{{Sierra} et~al.}{2024}]{sierra2024}
{Sierra} A.,  et~al., 2024, \mn@doi [\apj] {10.3847/1538-4357/ad7460}, \href {https://ui.adsabs.harvard.edu/abs/2024ApJ...974..306S} {974, 306}

\bibitem[\protect\citeauthoryear{{Stadler}, {G{\'a}rate}, {Pinilla}, {Lenz}, {Dullemond}, {Birnstiel}  \& {Stammler}}{{Stadler} et~al.}{2022}]{stadler2022}
{Stadler} J.,  {G{\'a}rate} M.,  {Pinilla} P.,  {Lenz} C.,  {Dullemond} C.~P.,  {Birnstiel} T.,   {Stammler} S.~M.,  2022, \mn@doi [\aap] {10.1051/0004-6361/202243338}, \href {https://ui.adsabs.harvard.edu/abs/2022A&A...668A.104S} {668, A104}

\bibitem[\protect\citeauthoryear{{Stammler} \& {Birnstiel}}{{Stammler} \& {Birnstiel}}{2022}]{stammler2022}
{Stammler} S.~M.,  {Birnstiel} T.,  2022, \mn@doi [\apj] {10.3847/1538-4357/ac7d58}, \href {https://ui.adsabs.harvard.edu/abs/2022ApJ...935...35S} {935, 35}

\bibitem[\protect\citeauthoryear{{Tazzari}, {Beaujean}  \& {Testi}}{{Tazzari} et~al.}{2018}]{tazarri2018}
{Tazzari} M.,  {Beaujean} F.,   {Testi} L.,  2018, \mn@doi [\mnras] {10.1093/mnras/sty409}, \href {https://ui.adsabs.harvard.edu/abs/2018MNRAS.476.4527T} {476, 4527}

\bibitem[\protect\citeauthoryear{{Toci}, {Rosotti}, {Lodato}, {Testi}  \& {Trapman}}{{Toci} et~al.}{2021}]{toci2021}
{Toci} C.,  {Rosotti} G.,  {Lodato} G.,  {Testi} L.,   {Trapman} L.,  2021, \mn@doi [\mnras] {10.1093/mnras/stab2112}, \href {https://ui.adsabs.harvard.edu/abs/2021MNRAS.507..818T} {507, 818}

\bibitem[\protect\citeauthoryear{{Trapman}, {Facchini}, {Hogerheijde}, {van Dishoeck}  \& {Bruderer}}{{Trapman} et~al.}{2019}]{trapman2019}
{Trapman} L.,  {Facchini} S.,  {Hogerheijde} M.~R.,  {van Dishoeck} E.~F.,   {Bruderer} S.,  2019, \mn@doi [\aap] {10.1051/0004-6361/201834723}, \href {https://ui.adsabs.harvard.edu/abs/2019A&A...629A..79T} {629, A79}

\bibitem[\protect\citeauthoryear{{Trapman}, {Rosotti}, {Zhang}  \& {Tabone}}{{Trapman} et~al.}{2023}]{trapman2023}
{Trapman} L.,  {Rosotti} G.,  {Zhang} K.,   {Tabone} B.,  2023, \mn@doi [\apj] {10.3847/1538-4357/ace7d1}, \href {https://ui.adsabs.harvard.edu/abs/2023ApJ...954...41T} {954, 41}

\bibitem[\protect\citeauthoryear{{Trapman} et~al.,}{{Trapman} et~al.}{2025a}]{trapman2025_mass}
{Trapman} L.,  et~al., 2025a, \mn@doi [arXiv e-prints] {10.48550/arXiv.2506.10738}, \href {https://ui.adsabs.harvard.edu/abs/2025arXiv250610738T} {p. arXiv:2506.10738}

\bibitem[\protect\citeauthoryear{{Trapman} et~al.,}{{Trapman} et~al.}{2025b}]{trapman2025_size}
{Trapman} L.,  et~al., 2025b, \mn@doi [arXiv e-prints] {10.48550/arXiv.2506.10750}, \href {https://ui.adsabs.harvard.edu/abs/2025arXiv250610750T} {p. arXiv:2506.10750}

\bibitem[\protect\citeauthoryear{{Tripathi}, {Andrews}, {Birnstiel}  \& {Wilner}}{{Tripathi} et~al.}{2017}]{tripathi2017}
{Tripathi} A.,  {Andrews} S.~M.,  {Birnstiel} T.,   {Wilner} D.~J.,  2017, \mn@doi [\apj] {10.3847/1538-4357/aa7c62}, \href {https://ui.adsabs.harvard.edu/abs/2017ApJ...845...44T} {845, 44}

\bibitem[\protect\citeauthoryear{{Valencia}, {Moro-Martin}  \& {Teske}}{{Valencia} et~al.}{2025}]{valencia2025}
{Valencia} D.,  {Moro-Martin} A.,   {Teske} J.,  2025, \mn@doi [arXiv e-prints] {10.48550/arXiv.2505.09754}, \href {https://ui.adsabs.harvard.edu/abs/2025arXiv250509754V} {p. arXiv:2505.09754}

\bibitem[\protect\citeauthoryear{{Vioque} et~al.,}{{Vioque} et~al.}{2025}]{vioque2025_age-pro}
{Vioque} M.,  et~al., 2025, \mn@doi [arXiv e-prints] {10.48550/arXiv.2506.10746}, \href {https://ui.adsabs.harvard.edu/abs/2025arXiv250610746V} {p. arXiv:2506.10746}

\bibitem[\protect\citeauthoryear{Virtanen et~al.,}{Virtanen et~al.}{2020}]{SciPy_2020}
Virtanen P.,  et~al., 2020, \mn@doi [Nature Methods] {10.1038/s41592-019-0686-2}, 17, 261

\bibitem[\protect\citeauthoryear{{Winn} \& {Fabrycky}}{{Winn} \& {Fabrycky}}{2015}]{winn2015}
{Winn} J.~N.,  {Fabrycky} D.~C.,  2015, \mn@doi [\araa] {10.1146/annurev-astro-082214-122246}, \href {https://ui.adsabs.harvard.edu/abs/2015ARA&A..53..409W} {53, 409}

\bibitem[\protect\citeauthoryear{{Zhang}, {Kalscheur}, {Long}, {Zhang}, {Long}, {Bergin}, {Zhu}  \& {Trapman}}{{Zhang} et~al.}{2023}]{zhang_s2023}
{Zhang} S.,  {Kalscheur} M.,  {Long} F.,  {Zhang} K.,  {Long} D.~E.,  {Bergin} E.~A.,  {Zhu} Z.,   {Trapman} L.,  2023, \mn@doi [\apj] {10.3847/1538-4357/acd334}, \href {https://ui.adsabs.harvard.edu/abs/2023ApJ...952..108Z} {952, 108}

\bibitem[\protect\citeauthoryear{{Zhang} et~al.,}{{Zhang} et~al.}{2025}]{zhang_age-pro2025}
{Zhang} K.,  et~al., 2025, \mn@doi [arXiv e-prints] {10.48550/arXiv.2506.10719}, \href {https://ui.adsabs.harvard.edu/abs/2025arXiv250610719Z} {p. arXiv:2506.10719}

\bibitem[\protect\citeauthoryear{{Zormpas}, {Birnstiel}, {Rosotti}  \& {Andrews}}{{Zormpas} et~al.}{2022}]{zormpas2022}
{Zormpas} A.,  {Birnstiel} T.,  {Rosotti} G.~P.,   {Andrews} S.~M.,  2022, \mn@doi [\aap] {10.1051/0004-6361/202142046}, \href {https://ui.adsabs.harvard.edu/abs/2022A&A...661A..66Z} {661, A66}

\bibitem[\protect\citeauthoryear{{van Terwisga} \& {Hacar}}{{van Terwisga} \& {Hacar}}{2023}]{vanTerwisga2023}
{van Terwisga} S.~E.,  {Hacar} A.,  2023, \mn@doi [\aap] {10.1051/0004-6361/202346135}, \href {https://ui.adsabs.harvard.edu/abs/2023A&A...673L...2V} {673, L2}

\bibitem[\protect\citeauthoryear{{van Terwisga} et~al.,}{{van Terwisga} et~al.}{2020}]{vanTerwisga2020}
{van Terwisga} S.~E.,  et~al., 2020, \mn@doi [\aap] {10.1051/0004-6361/201937403}, \href {https://ui.adsabs.harvard.edu/abs/2020A&A...640A..27V} {640, A27}

\bibitem[\protect\citeauthoryear{{van der Marel} \& {Pinilla}}{{van der Marel} \& {Pinilla}}{2023}]{vanderMarel2023}
{van der Marel} N.,  {Pinilla} P.,  2023, \mn@doi [arXiv e-prints] {10.48550/arXiv.2310.09077}, \href {https://ui.adsabs.harvard.edu/abs/2023arXiv231009077V} {p. arXiv:2310.09077}

\makeatother
\end{thebibliography}

%%%%%%%%%%%%%%%%%%%%%%%%%%%%%%%%%%%%%%%%%%%%%%%%%%

%%%%%%%%%%%%%%%%% APPENDICES %%%%%%%%%%%%%%%%%%%%%

\appendix
\section{Tables with disc properties}

Tables~\ref{table:resolved_discs} and ~\ref{table:unresolved_discs} summarise the main properties of the resolved and unresolved discs in UpperSco analysed in this study. The complete machine-readable tables are available in \texttt{zenodo} at \url{https://doi.org/10.5281/zenodo.17182603}.

\begin{table*}
\centering
\caption{Properties of resolved discs.  The complete machine-readable table is available at \url{https://doi.org/10.5281/zenodo.17182603}.}
\label{table:resolved_discs}
\begin{sideways}
    \begin{tabular}{|c|c|c|c|c|c|c|c|c|c|c|c|c|c|}
    \hline
        Name & Group  Name & SpT & d [pc] & $M_\star$ [$M_\odot$]& $L_\star$ [$L_\odot$] & $F_{\rm{UV}}$ [$G_0$] & dRa [$^{\prime \prime}$]& dDec [$^{\prime \prime}$]& $i$[$^{\circ}$] & PA[$^{\circ}$] & Flux  [mJy] & $R_{90}$ [$^{\prime \prime}$] & $R_{68}$ [$^{\prime \prime}$] \\ \hline
        \small{2MASS J16042165 2130284} & delta sco & K2 & 144.6 & $1.05\pm{0.07}$  & 0.52 & $24.56^{+11.16}_{-9.55}$ & $0.20$ & $0.41$ & $11.90^{+0.15}_{-0.18}$ &$141.34^{+0.44}_{-0.72}$ & $223.30^{+0.52}_{-0.45}$ & $0.766^{+0.001}_{-0.001}$ & $0.672^{+0.001}_{-0.001}$ \\ \hline
    \end{tabular}
    \end{sideways}
\end{table*}

\begin{table*}
\centering
\caption{Properties of the  unresolved discs.  The complete machine-readable table is available at \url{https://doi.org/10.5281/zenodo.17182603}.}
\label{table:unresolved_discs}

    \begin{tabular}{|c|c|c|c|c|c|c|c|c|c|}
    \hline
        Name & Group  Name & SpT & d [pc] & $M_\star$ [$M_\odot$]& $L_\star$ [$L_\odot$] & $F_{\rm{UV}}$ [$G_0$] & dRa [$^{\prime \prime}$]& dDec [$^{\prime \prime}$]&  Flux  [mJy]  \\ \hline
        2MASS\,J15540240\,2254587 & delta sco & M4.5  & $141.9$ & $0.11\pm{0.06}$ & $0.03$& $9.97^{+1.51}_{-1.31}$&$-0.03^{+0.02}_{-0.01}$&$-0.02^{+0.01}_{-0.01}$& $1.95\pm{0.2}$  \\ \hline
    \end{tabular}
\end{table*}

\section{Complementary Figures}
\label{app:comparison}

In this appendix, we include several complementary figures to the main paper. 

\begin{itemize}

\item Figure~\ref{fig:Comparison_Literature} compares the  the disc geometry, sizes, and fluxes in this work to values reported in \cite{carpenter2025} for the dust continuum emission (dots) and from \zagaria \,for CO emission (crosses).

\begin{figure*}
    \centering
    \includegraphics[width=\linewidth]{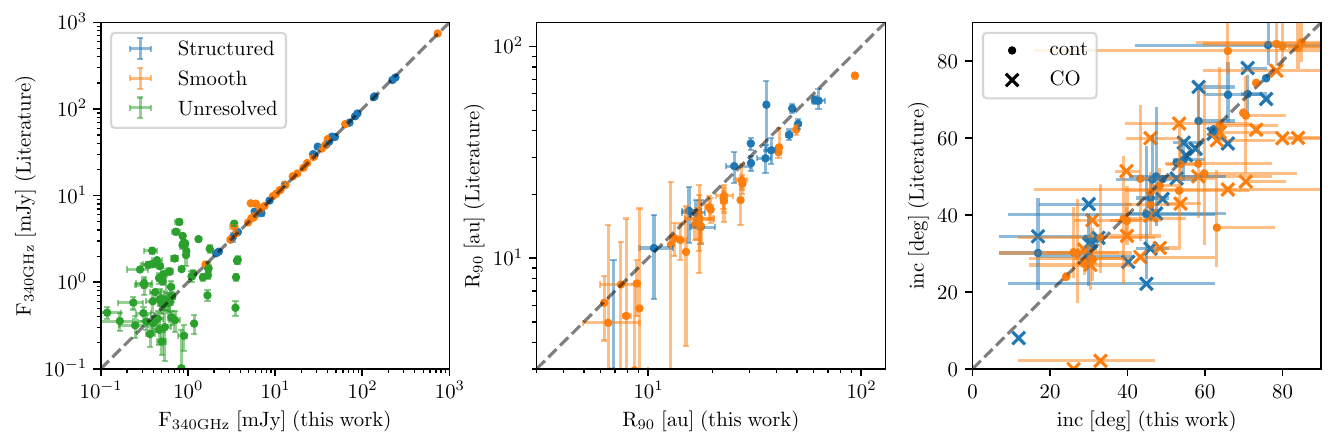}
    \caption{Comparison of the disc geometry, sizes, and fluxes in this work with literature values. From left to right: Dust continuum fluxes, dust continuum radius ($R_{90}$),  and  disc inclination. Dot points come from the values obtained from the continuum observations reported in Carpenter et al. (2025), while crosses are from CO estimations from Zagaria et al. (in prep).}
    \label{fig:Comparison_Literature}
\end{figure*}

%%%%%%%%%%%%%%%%%%%%%%%%%%%%%%%%%%%%%%%%%%%%%%%%%%

\item Figure~\ref{fig:dust_density_1Msun} shows the dust density distribution after 1\,Myr of evolution for discs around a 1.0\,$M_\odot$, assuming $F_{\rm{UV}}=10G_0$ and different type of pressure traps in the models.

\begin{figure*}
\centering
\includegraphics[width=17cm]{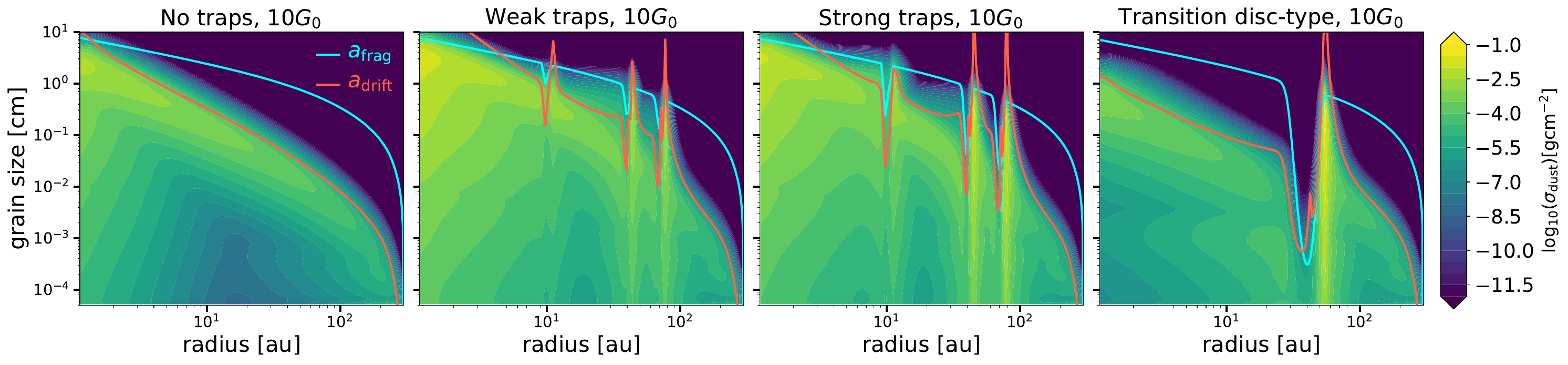}
\caption{Dust density distribution after 2\,Myr of evolution for discs with (from left to right): no traps, weak traps, strong-traps, and transition disc-type trap. All of these models assume a $F_{\rm{UV}}=10\,G_0$. These models are for the case of discs around a 1.0\,$M_\odot$.}
\label{fig:dust_density_1Msun}
\end{figure*}

\item Figure~\ref{properties_observables_1Msun} shows the evolution of the millimetre flux, $R_{90}$, $R_{\rm{CO}}$ and $R_{\rm{CO}}/R_{90}$ for different values of $F_{\rm{UV}}$, and for different pressure traps. These results correspond to the models of discs around a 1.0\,$M_\odot$ star.

\begin{figure*}
\centering
\includegraphics[width=17cm]{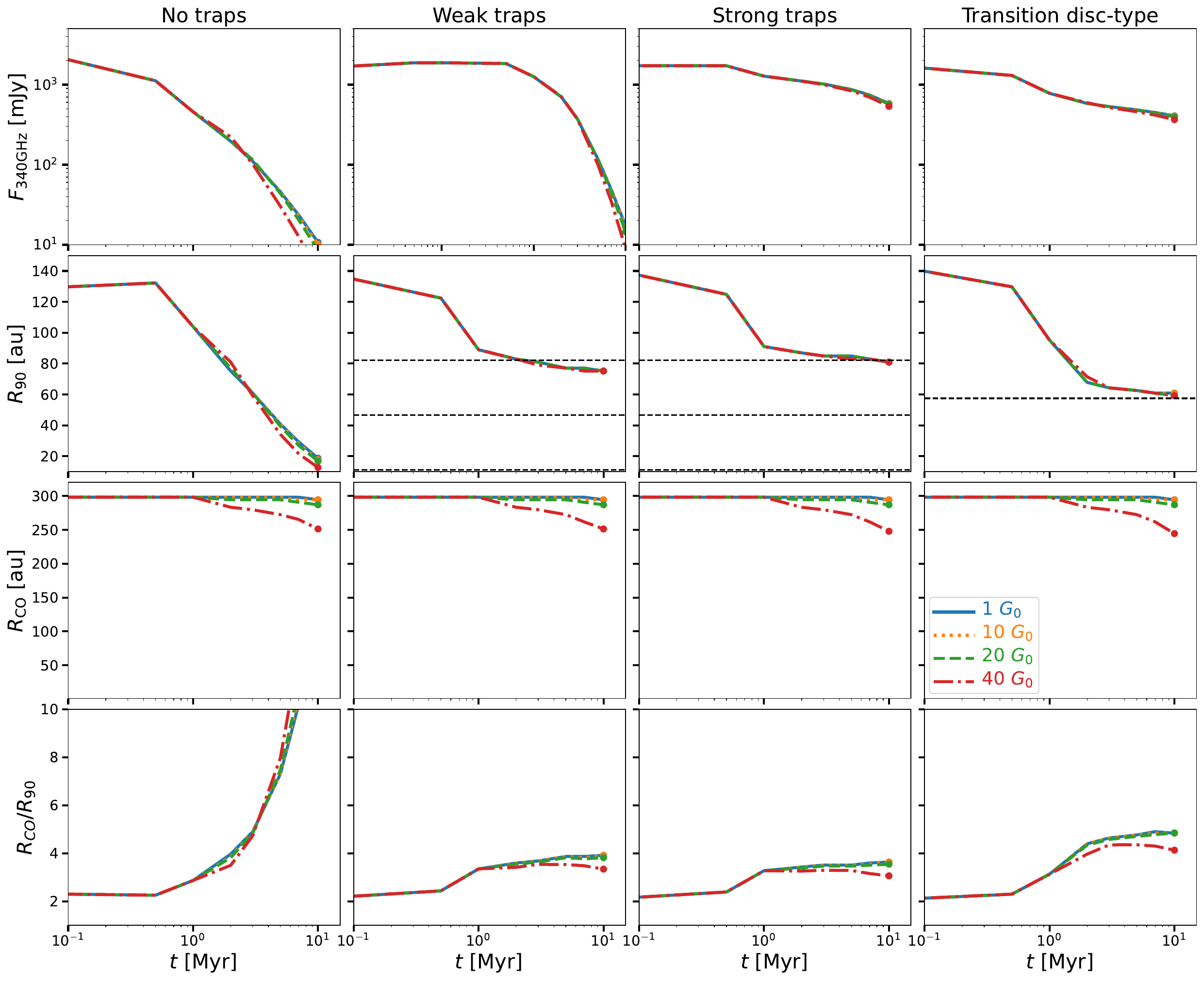}
\caption{From top to bottom: evolution of the millimeter flux, $R_{90}$, $R_{\rm{CO}}$ and $R_{\rm{CO}}/R_{90}$ for different values of $G_0$. From left to right: no traps, weak traps, strong-traps, and transition disc-type trap. These results correspond to the models of discs around a 1.0\,$M_\odot$ star.}
\label{properties_observables_1Msun}
\end{figure*}

\end{itemize}

\bsp	% typesetting comment
\label{lastpage}
\end{document}